\documentclass{aa} % [letter] for the letters 
\usepackage{graphicx}
\usepackage{txfonts}
\usepackage{multirow}
\usepackage{colortbl}

\usepackage{hyperref}  
\hypersetup{colorlinks=true,linkcolor=[rgb]{1.,0.2,0.2},citecolor=[rgb]{0.1,0.4,1.},filecolor=[rgb]{0.7,0.2,0.2},urlcolor=[rgb]{0.7,0.2,0.2}}
\usepackage{color}
\definecolor{blue}{rgb}{0., 0., 1}

\newcommand{\oiii}{[\textrm{O}\textsc{iii}]}

\newcommand{\oiiiv}{[\textrm{O}\textsc{iii}]\ensuremath{\lambda5007}}

\newcommand{\ha}{\ifmmode {\rm H}\alpha \else H$\alpha$\fi}
\newcommand{\halam}{\ifmmode {\rm H}\alpha \lambda6563 \else H$\alpha$ $\lambda$6563 \fi}
\newcommand{\hb}{\ifmmode {\rm H}\beta \else H$\beta$\fi}
\newcommand{\hg}{\ifmmode {\rm H}\gamma \else H$\gamma$\fi}
\newcommand{\hblam}{\ifmmode {\rm H}\beta \lambda4861 \else H$\beta$ $\lambda$4861 \fi}
\newcommand{\lya}{\ifmmode {\rm Ly}\alpha \else Ly$\alpha$\fi}
\newcommand{\pg}{\ifmmode {\rm P}\gamma \else Pa$\gamma$\fi}
\newcommand{\lyb}{\ifmmode {\rm Ly}\beta \else Ly$\beta$\fi}
\newcommand{\lyg}{\ifmmode {\rm Ly}\gamma \else Ly$\gamma$\fi}

\newcommand{\flyc}{\ifmmode  \mathrm{f}_\mathrm{esc}\mathrm{(LyC)} \else $\mathrm{f}_\mathrm{esc}\mathrm{(LyC)}$\fi}

\def\ergs{\ifmmode \mathrm{erg\hspace{1mm}s}^{-1} \else erg s$^{-1}$\fi}

\def\micron{\ifmmode \mu\mathrm{m} \else $\mu$m\fi}
\def\msun{\ifmmode \mathrm{M}_{\odot} \else M$_{\odot}$\fi}
\def\msunyr{\ifmmode \mathrm{M}_{\odot} \hspace{1mm}{\rm yr}^{-1} \else $\mathrm{M}_{\odot}$ yr$^{-1}$\fi}
\def\zsun{\ifmmode Z_{\odot} \else Z$_{\odot}$\fi}
\def\lsun{\ifmmode L_{\odot} \else L$_{\odot}$\fi}
\def\mstar{\ifmmode \mathrm{M}_{\star} \else M$_{\star}$\fi}

\newcommand\TM{M25}
\defcitealias{Morishita2025}{\TM}

\usepackage{orcidlink}

\newcommand{\orcid}[1]{\href{https://orcid.org/#1}{\textcolor[HTML]{A6CE39}{\aiOrcid}}}

\begin{document}

\titlerunning{chemical inhomogeneity at 10-parsec scales in LAP2}

\title{A pocket of oxygen in a near-pristine galaxy at z=4.2: chemical inhomogeneity at 10-parsec scales in LAP2}

\authorrunning{Matteo Messa et al.}
\author{
M.~Messa\inst{\ref{inafbo}}\fnmsep\thanks{E-mail: \href{mailto:matteo.messa@inaf.it}{matteo.messa@inaf.it}}$^{\orcidlink{0000-0003-1427-2456}}$ \and
E.~Vanzella\inst{\ref{inafbo}}$^{\orcidlink{0000-0002-5057-135X}}$ \and
S.~Rojas-Ruiz\inst{\ref{UCLA}}$^{\orcidlink{0000-0003-2349-9310}}$ \and
T.~Treu\inst{\ref{UCLA}}$^{\orcidlink{0000-0002-8460-0390}}$ \and
P.~Bergamini\inst{\ref{inafbo}}$^{\orcidlink{0000-0003-1383-9414}}$\and
G.~Di~Rosa \inst{\ref{unife},\ref{inafbo}}$^{\orcidlink{0000-0002-6813-0632}}$\and
M.~Lombardi\inst{\ref{unimi}}$^{\orcidlink{0000-0002-3336-4965}}$ \and
P.~Rosati \inst{\ref{unife},\ref{inafbo}}$^{\orcidlink{0000-0002-6813-0632}}$\and
A.~Zanella\inst{\ref{inafbo}}$^{\orcidlink{0000-0001-8600-7008}}$\and
A.~Adamo\inst{\ref{univstock}}$^{\orcidlink{0000-0002-8192-8091}}$\and
A.~Bolamperti\inst{\ref{mpa},\ref{inafiasf}}$^{\orcidlink{0000-0001-5976-9728}}$\and
F.~Calura\inst{\ref{inafbo}}$^{\orcidlink{0000-0002-6175-0871}}$\and
M.~Castellano\inst{\ref{inafroma}}$^{\orcidlink{0000-0001-9875-8263}}$\and
R.~Gilli\inst{\ref{inafbo}}$^{\orcidlink{0000-0001-8121-6177}}$\and
C.~Grillo\inst{\ref{unimi},\ref{inafiasf}}$^{\orcidlink{0000-0002-5926-7143}}$\and
U.~Meštrić\inst{\ref{ICRAR},\ref{unibelgr}}$^{\orcidlink{0000-0002-0441-8629}}$
}
\institute{
INAF -- OAS, Osservatorio di Astrofisica e Scienza dello Spazio di Bologna, via Gobetti 93/3, I-40129 Bologna, Italy \label{inafbo} 
%\and
%IPAC, California Institute of Technology, Pasadena, CA 91125, USA \label{ipac} 
%\and
%Astronomical Institute, Tohoku University, 6-3 Aramaki, Aoba-ku, Sendai 980-8578, Japan\label{tohoku}
%\and 
%Space Telescope Science Institute (STScI), 3700 San Martin Drive, Baltimore, MD 21218, USA\label{stsci}
\and
Department of Physics and Astronomy, University of California, Los Angeles, 430 Portola Plaza, Los Angeles, CA 90095, USA\label{UCLA}
\and
Dipartimento di Fisica e Scienze della Terra, Università degli Studi di Ferrara, Via Saragat 1, I-44122 Ferrara, Italy\label{unife}
\and
Dipartimento di Fisica, Università degli Studi di Milano, Via Celoria 16, I-20133 Milano, Italy\label{unimi}
\and
Department of Astronomy, Oskar Klein Centre, Stockholm University, AlbaNova University Center, SE-106 91, Sweden\label{univstock}
%\and
%MIT Kavli Institute for Astrophysics and Space Research, 70 Vassar Street, Cambridge, MA 02139, USA\label{mitkavli}
%\and
%Astronomical Institute, Graduate School of Science, Tohoku University, Sendai, Miyagi 980-8578, Japan\label{gradsc_tohoku}
\and
Max-Planck-Institut f\"ur Astrophysik, Karl-Schwarzschild-Str. 1, D-85748 Garching, Germany\label{mpa}
%\and
%Univ Lyon, Univ Lyon1, ENS de Lyon, CNRS, Centre de Recherche Astrophysique de Lyon UMR5574, Saint-Genis-Laval, France\label{CRAL}
%\and
%Observatoire de Genève, Université de Genève, Chemin Pegasi 51, 1290 Versoix, Switzerland\label{obsgeneve}
\and 
INAF -- IASF Milano, via A. Corti 12, I-20133 Milano, Italy\label{inafiasf}
\and 
INAF -- Osservatorio Astronomico di Roma, Via Frascati 33, 00078 Monteporzio Catone, Rome, Italy\label{inafroma}
\and
ICRAR, The University of Western Australia, 35 Stirling Highway, Crawley, WA 6009, Australia\label{ICRAR}
\and
Department of Astronomy, Faculty of Mathematics, University of Belgrade, Studentski trg 16, 11000 Belgrade, Serbia\label{unibelgr}
}

\date{} %Received June 14, 2020; accepted July 14, 2020}
 
\abstract
{Extremely metal-poor galaxies (EMPGs) at high redshift trace the earliest stages of chemical enrichment and are the closest observable counterparts of the environments in which Population~III (Pop~III) stars may form. The faintest of these systems are accessible in detail only when strongly magnified by gravitational lensing.}
{We aim to constrain the oxygen abundance, the ionizing properties, and the spatial structure of LAP2, a strongly lensed (with a magnification $\mu\simeq50$), oxygen-deficient galaxy at $z=4.2$ behind the galaxy cluster Abell~2744.}
{We present new JWST/NIRSpec IFU PRISM observations ($\sim$13.6~hr) of LAP2, which we combine with JWST/NIRCam imaging to map the H$\alpha$, H$\beta$, and [O\,{\sc iii}]$\lambda$5007 emission, and to model the morphology and the spectral energy distribution of the system.}
{LAP2 is intrinsically extremely faint ($M_{\rm UV}=-12.7$) and has a low stellar mass ($\rm M_\star\sim 10^{5}\,M_\odot$), comparable to that of massive star clusters. It consists of a compact core ($\rm R_{\rm eff}\sim7$~pc) embedded in a more diffuse emission (with $\rm R_{\rm eff}\sim20$~pc). Strong H$\alpha$ emission (EW$_{\rm rest}\simeq880$~\AA, $\log\xi_{\rm ion}\simeq25.3$), a blue UV slope ($\beta_{\rm UV}\simeq-2.5$, down to $\simeq-3.1$ in the core), and a Balmer decrement consistent with no dust attenuation indicate a young ($\lesssim10$~Myr), dust-free burst. [O\,{\sc iii}]$\lambda$5007 is not detected in the integrated spectrum, implying [O\,{\sc iii}]/H$\beta\leq0.61$ ($3\sigma$) and, using an extrapolated strong-line calibration, $12+\log({\rm O/H})\leq6.4$ ($\lesssim0.5\%\,Z_\odot$). This places LAP2 among the most metal-poor galaxies known. Faint [O\,{\sc iii}] emission is, however, detected in the core (with $\rm4\sigma$ significance), where we derive $12+\log({\rm O/H})\simeq6.6$ ($\sim1\%\,Z_\odot$). This difference can be explained either by local self-enrichment, requiring only $\rm \sim0.2\,M_\odot$ of newly synthesized oxygen, or by a higher ionization parameter in the core. The upper limit on He\,{\sc ii}$\lambda$1640 (EW$_{\rm rest}<39$~\AA, $3\sigma$) does not exclude a contribution from metal-free stars. We also find a tentative H$\alpha$-only component offset by $\gtrsim30$~pc from the stellar emission, possibly tracing gas ionized by escaping radiation.}
{LAP2 extends the census of spectroscopically confirmed EMPGs to star-cluster masses and to $M_{\rm UV}\approx-13$, and it lies below the [O\,{\sc iii}]/H$\beta$ values typical of faint high-redshift galaxies. The spatially resolved data show a localized inhomogeneous chemical enrichment at $\sim10$~pc scales. Its location in a low-density region, in a redshift gap between two galaxy concentrations, supports a scenario in which underdense environments preserve nearly pristine gas well after reionization.}
% OPTIONAL (once Sect. 4.5 is final): add before the last sentence
% "LAP2 resides in a sparsely populated region, [with no spectroscopically confirmed neighbour within ~NUM pMpc], supporting a scenario in which low-density environments preserve nearly pristine gas well after reionization."
   \maketitle

\section{Introduction}
\label{sect:intro}

The metallicity of galaxies is closely linked to their stellar mass, giving rise to the well-established mass–metallicity relation (MZR, e.g. \citealp{Garnett2002,Tremonti2004}). This relation has been confirmed over a broad range of stellar masses, including the dwarf-galaxy regime \citep[e.g.,][]{Lee2006,Andrews_Martini2013} establishing it as the fundamental scaling relation describing the chemical evolution of galaxies.
The physical origin of the MZR is generally attributed to the interplay between the production of metals by stars, the accretion of relatively metal-poor gas, and the removal of metal-enriched material through galactic outflows \citep[e.g.,][]{Larson1974,Dave2012,Maiolino_Mannucci2019}.
The dependence of metallicity on star-formation activity at fixed stellar mass led to the formulation of a more fundamental metallicity relation (FMR), providing a framework in which the evolution of the MZR can be interpreted in terms of the changing gas accretion and star-formation properties of galaxies across cosmic time \citep[e.g.][]{Mannucci2010,LaraLopez2010}.
%in which stellar mass, star-formation rate, and gas-phase metallicity form a three-dimensional scaling relation \citep{Mannucci2010}. The FMR 

The evolution of the MZR with cosmic time was established progressively through observations of star-forming galaxies at high redshift, initially reaching $\rm z\sim3-4$ \citep[e.g.][]{Erb2006,Maiolino2008,Mannucci2009,Kewley_Ellison2008}. The advent of the James Webb Space Telescope (JWST), with its unprecedented sensitivity and wavelength coverage, provided access to the rest-frame optical emission lines over the entire redshift range up to $\rm z\sim10$. 
%Early JWST studies already extended metallicity measurements into the epoch of early galaxy assembly, finding a relatively mild evolution of the MZR (and FMR) from $z\approx2-3$, to $\rm z>4$, with high-z galaxies being systematically more metal-poor than what expected from their stellar masses and star-formation rates \citep[e.g.][]{nakajima2023,Curti2024,He2026,Hsiao2026_arXiv,Lam2026_arXiv}.
Early JWST studies already extended metallicity measurements into the epoch of early galaxy assembly, finding only a mild decrease in the normalization of the MZR from $z\approx2-3$ to $z>4$ \citep[e.g.][]{nakajima2023,Curti2024,He2026,Hsiao2026_arXiv,Lam2026_arXiv}.
When star-formation rates are also taken into account, several studies find that galaxies at $z\gtrsim6$ are, on average, more metal-poor than predicted by the FMR calibrated on local galaxies, by up to $\sim0.5$~dex \citep[e.g.][]{Curti2024}, possibly indicating dilution by accretion of pristine gas \citep[e.g.][]{Heintz2023}. However, this offset is not universal,and some high-redshift galaxies lie on the local relations \citep[e.g.,][]{PerezDiaz2026}.

The low-mass and low-metallicity end of the MZR is particularly relevant for understanding the earliest stages of galaxy evolution. Extremely metal-poor galaxies (EMPGs), generally defined as systems with $12+\log({\rm O/H})\lesssim7.3$ are rare in the local Universe. They are thought to result from a combination of inefficient metal retention, metal-poor gas accretion, and bursty star-formation histories \citep[e.g.][]{Izotov1999,Thuan2005,Lee2006,Kojima2020}.
At high redshift, the low-mass end of the MZR is expected to extend towards even lower metallicities than in the local universe. Recent JWST observations of faint galaxies have begun to reveal this population, with measurements reaching estimated metallicities of a few per cent of the solar value and, in some cases, even lower \citep[e.g.][]{Vanzella2023_lap1,Vanzella2026b,Maiolino2024,Maiolino2026,Maiolino2026_arXiv,Nakajima2025,Hsiao2025,Morishita2025,Cai2025,Ubler2026,Li2026_arXiv}. 
Such low metallicities approach the regime expected for Population III (Pop III) stars, the first generation of stars formed from primordial metal-free gas \citep[see][for a recent review]{Venditti2026_review}. Their metal deficiency can be probed through the absence or extreme weakness of oxygen and other metal emission lines, as expected for nebular gas with primordial composition \citep[e.g.,][]{Zackrisson2011,Nakajima_Maiolino2022,Trussler2023,Fujimoto2025_arxiv}. A complementary signature is strong He~\textsc{II} emission, particularly the He II \(\lambda1640\) recombination line, which arises from the exceptionally hard ionizing spectra predicted for hot, massive metal-free stars \citep[e.g.,][]{schaerer2002,schaerer2003,Raiter2010}. High He II equivalent widths can therefore provide a sensitive probe of a very hard, potentially primordial ionizing population.
%although such emission is not uniquely produced by Pop III stars and can also arise from other hard ionizing sources.
%Such low metallicities start approaching the expected ranges for Population III (Pop III) stars, the first generation of stars formed from primordial, metal-free gas. Their metal deficiency is inferred from the lack of oxygen emission lines, ideally tracing the putative pristine composition of the interstellar medium (ISM) surrounding them \citep[e.g.,][]{Zackrisson2011,Nakajima_Maiolino2022,Trussler2023}, or by large equivalent widths (EWs) of the He II line, directly linked to the hard ionizing spectrum of a high-mass, metal-free star (References).

The detection of Pop III stars, one of the most sought-after discoveries in modern astrophysics, faces multiple challenges. Due to the extremely short lifetimes of Pop III stars, their main spectroscopic features fade rapidly, already $\sim3$ Myr after the initial burst, making them statistically rare to observe \citep[e.g.,][]{Raiter2010,Zackrisson2011}. This timescale, however, primarily characterizes the rapid fading of the hard ionizing spectrum following an instantaneous burst; more extended star-formation histories can sustain nebular emission for longer periods, while cooling radiation and supernova remnants may provide additional signatures after the most massive Pop III stars have evolved \citep[e.g.][]{schaerer2002,schaerer2003,Trussler2023}.
Theoretical models expect Pop III stars to form in pristine low-mass halos (minihalos and atomic-cooling halos), or in the faint outskirts of more massive halos at high redshift that are still accreting pristine gas \citep[e.g.][]{abel2002,Yoshida2003,Xu2013,Venditti2023,Venditti2024,Venditti2025,Koller2026,Storck2026}, resulting in faint continuum and line emission \citep{zack2015}. 
Because of this, many of the currently known Pop III candidates and EMPGs are gravitationally lensed systems, where the observed fluxes are magnified by factors of tens to hundreds, e.g. LAP2 at $z=4.2$ \citep{Vanzella2026b}, AMORE6 at $\rm z=5.7$ \citep{Morishita2025}, LAP1 at $\rm z=6.6$ \citep{Vanzella2023_lap1,Nakajima2025} and Hebe at $\rm z=10.6$ \citep{Maiolino2024,Maiolino2026_arXiv,Ubler2026}. Interestingly, some of these are observed at relatively low redshifts, $z\lesssim6$ (down to $z\approx3$ in the case of the non-lensed system CR3, \citealp{Cai2025}), suggesting that Pop III systems may form later than initially predicted, at least in some locations. Due to the spatially inhomogeneous nature of cosmic metal enrichment, pockets of pristine gas may survive until relatively low redshifts, allowing Population III star formation to persist well beyond the epoch of reionization, as shown by simulations \citep[e.g.,][]{Tornatore2007,Trenti2009a,Trenti2009b,Pallottini2014,Xu2016,Liu2020,Venditti2023,Storck2026,Venditti2026_review}.

Galaxies with metallicities approaching $\rm Z\lesssim1\%~Z_\odot$ are expected to be increasingly difficult to distinguish from systems hosting genuinely metal-free stellar populations, even when JWST spectroscopy is available. Part of the difficulty is the large uncertainty in the calibrations used to convert strong line ratios to actual oxygen abundances, especially because their low-metallicity end ($\rm \lesssim5\%~Z_\odot$) is blindly extrapolated from higher metallicity data. 
In addition, PopIII stars are expected to co-exist along with (or within) enriched environments/galaxies \citep[e.g.][]{Venditti2023,Goswami2026}.
For these reasons, EMPGs represent an important intermediate regime in the search for primordial stellar populations. Although a gas-phase metallicity of a few per cent of the solar value does not imply the presence of metal-free stars, such systems provide a window into the physical conditions under which the first generations of stars may have formed and can potentially preserve signatures of recent or incomplete chemical enrichment. 

In this context, the lensed galaxy LAP2 at $z=4.2$ stands out as one of the most oxygen-deficient systems currently known, based on the non-detection of strong oxygen lines (resulting in a \oiiiv/\hb\ upper limit $<0.85$) from NIRSpec/MSA observations \citep{Vanzella2026b}. LAP2 is a very faint galaxy, strongly lensed by the galaxy cluster Abell~2744, and was initially detected as a \lya\ emitting galaxy in the MUSE Lensing Cluster project \citep{Richard2021}.
In this work, we analyze new NIRSpec IFS observations of LAP2 designed to obtain stringent constraints on its oxygen deficiency, adding also a spatial and morphological analysis.
Unlike the previous MSA observations, which sampled only part of LAP2b, the IFU data cover the full extent of both counter-images, allowing us to measure the integrated emission without slit losses and to search for spatial variations of the metal content within the system.

The paper is organized as follows: in Section~\ref{sec:data_reduction}, the JWST observations are presented alongside their data reduction; the reduced data are analyzed in Section~\ref{sec:analysis_results}, where the main results are presented. The results are further discussed, in the context of the literature, in Section~\ref{sec:discussion}. We assume a flat $\Lambda$CDM cosmology with $\Omega_{\rm M}=0.315$ and $H_{0}=67.4\,{\rm km}\,{\rm s}^{-1}\,{\rm Mpc}^{-1}$ \citep{Planck2020}, the \citet{Kroupa2001} stellar initial mass function, and a solar abundance from \citet{asplund2009}, $\rm 12+log(O/H)_\odot = 8.69$. All magnitudes are in the AB system \citep{Oke_1983}, $m_{\rm AB}=23.9-2.5\log(f_\nu/\mu{\rm Jy})$. 
%In particular, \citet{Curti2024} found a shallower MZR slope at low stellar masses than inferred from more massive systems, while the increasing number of extremely metal-poor candidates identified with JWST is beginning to probe the regime where the relation may exhibit substantial intrinsic scatter. Understanding this population is crucial for determining whether the MZR represents a continuous sequence of chemical enrichment or whether the most metal-poor galaxies constitute a distinct population shaped by stochastic gas accretion and star formation.}
%

\section{Data Reduction}\label{sec:data_reduction}
\subsection{NIRSpec IFU}\label{sec:data_reduction:NIRSpec}
\begin{figure}
    \centering
    \includegraphics[width=0.37\columnwidth]{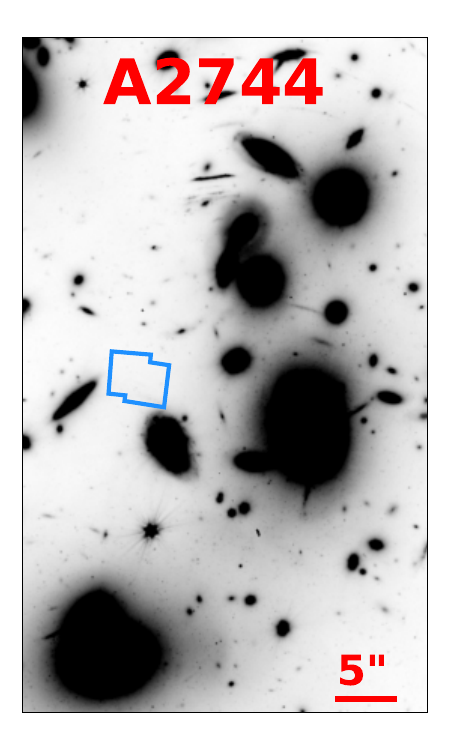}
    \includegraphics[width=0.62\columnwidth]{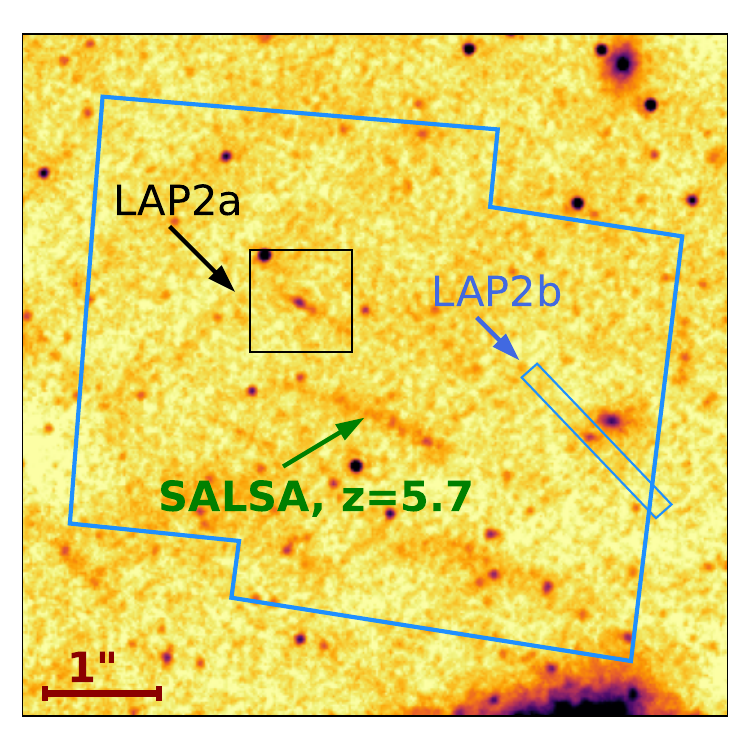}
    \includegraphics[width=0.99\columnwidth]{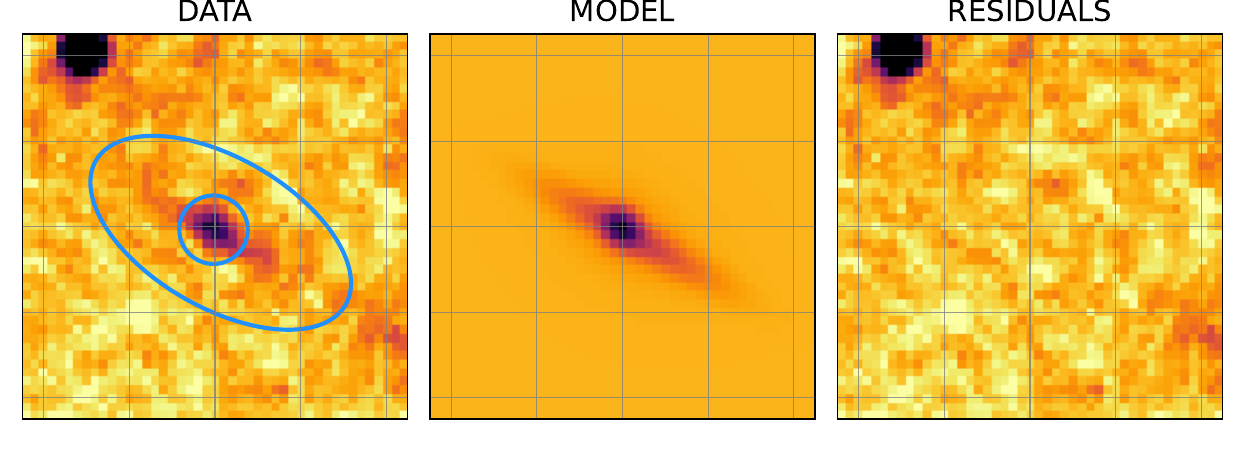}
    \caption{\textit{Top left:} Overview of the A2744 galaxy cluster, from a stack of the LW NIRCam filters. \textit{Top right:} zoom-in on the region of the LAP2 system, from a stack of F090W, F115W, F150W, and F200W, all PSF-matched to the F200W; the position of both counter-images `a' and `b' are shown (the latter along with the NIRSpec/MSA aperture from program GO 2561, analyzed in \citealp[]{Vanzella2026b}). In both top panels, the field-of-view (FoV) of the NIRSpec/IFU observations is over-plotted with a blue contour. A faint but highly magnified arc at $\rm z=5.66$, labeled `SALSA' (analyzed in \citealp{RojasRuiz2026_arxiv}), falls into the IFU FoV. The black square region on LAP2a is zoomed in the bottom panels. \textit{Bottom panels:} data, best-fit model and residuals of the LAP2a system; each square in the grid spans $\rm 0".2~arcsec$ by side. In the left panel, the apertures used for photometry (for the entire LAP2a and for its core region) are over-plotted. All panels are aligned with north-up and east-left.} 
    \label{fig:fov}
\end{figure}

The JWST/NIRSpec IFU observations used in this work (GO 7677, PI: Vanzella) targeted two galaxies, highly magnified by the galaxy cluster Abell 2744 (hereafter A2744): (1) the LAP2 system, at $\rm z=4.2$ (both counter-images `a' and `b', hereafter simply referred to as LAP2a and LAP2b) and (2) a $\rm Lyman-\alpha$ emitting galaxy (LAE) at $\rm z=5.66$, now recognized also in the stellar continuum as a very faint arc, named ``SALSA'' and analyzed in detail elsewhere \citep{RojasRuiz2026_arxiv}.
The data were acquired on October 25, 2025, and consist of PRISM spectroscopy in two adjacent pointings, covering the spectral range $\rm 0.8-5.3~\mu m$, with spectral resolution varying from $\rm R\simeq30$ (around $\rm \sim1~\mu m$) to $\rm R\simeq300$ (at the longest wavelengths). 
The effective exposure time per pointing was $6.8$ hr; LAP2a and the SALSA arclet were covered by both of the pointings, for a total exposure time of $13.62$ hr\footnote{LAP2a lies near the edge of the field of view in one pointing, so its signal comes mostly from the other; its effective depth is therefore lower than the nominal total.}. As LAP2 is strongly magnified (Section~\ref{sec:lens_model}), this exposure time corresponds to $\sim3.6\times10^4$~hr (i.e. $\sim4$ years) of integration to reach the same depth on an unlensed, unresolved source.
The layout of the field of view is shown in Fig.\ref{fig:fov}, both in the broader context of A2744 and in a zoomed-in region around the targets. 

The data were reduced following the same procedures described by \citet{Messa2025,Messa2026a}. We used the updated version of the STScI pipeline (\texttt{v3.0.0} and \texttt{jwst\_1584.pmap}); during stage 1 we performed the 1/f noise subtraction using the \texttt{clean\_flicker\_noise} step, and compare stage 1 products with and without this correction to assess its goodness. 
Although the pixels severely affected by light leakages from known failed open MSA shutters are already flagged by the pipeline, we noticed that automatic masking sometimes does not fully cover the leaking signal observed in the \texttt{rate} files. Thus, we visually build larger masks, checking all the exposure frames, and exclude the relative pixels from the further steps of the data reduction. In addition, we perform an aggressive flagging of all the pixels without robust detection (i.e. all pixels with DQ different from `0' or `4'), by excluding them after stage 2. 
The two pointings of the dataset are combined into a single mosaic during stage 3 of the pipeline. 
%For completeness, we also produced two individual cubes, one per pointing, and repeat the spectroscopic analysis of Section~\ref{sec:spectroscopy}, finding no significant differences (see Appendix~\ref{sec:app:spec_further}). 
For our analysis we used, as default, a cube reduced at 0.05 arcsec/pixel; a standard reduction at 0.1 arcsec/pixel was used for consistency check, finding no significant differences in the main results of this work (see Appendix~\ref{sec:app:spec_further}). 
In addition to the two main targets, when stacking the cube over several slices, we detect the continuum of other unresolved sources falling in the FoV; we used those to astrometrically align the cube to the NIRCam images. 

The sky background is estimated in each wavelength slice as the median estimated in a region that excludes the main targets (LAP2 and SALSA), the outskirts of a cluster galaxy at the bottom-right (see Fig.~\ref{fig:fov}) and all the bright point-like sources falling in the FoV. 
When analyzing the 1D spectra, other methods are tested, namely subtracting the background from a small mask around the target, or subtracting a running median of the signal spectrum (excluding the wavelengths covering the lines, see \citealp{Messa2025}); no major difference is found between different subtractions.
%We show in appendix~\ref{sec:app:spec_further} that the choice of the background estimator does not affect the metallicity measurement of LAP2.

When extracting the 1D spectra of the targets (see Section~\ref{sec:spectroscopy}) we noticed that the formal noise, resulting from the reduction pipeline, underestimates the observed scatter in the spectra, as previously reported in other NIRSpec observations \citep[e.g.,][]{ubler23,Fujimoto2025_ALPINE-CRISTAL}. For this reason, we decided to use the running standard deviation of the signal spectrum (over 100 wavelength slices, i.e. a range of $\rm \sim0.5~\mu m$) as the associated uncertainty.

\subsection{NIRCam}\label{sec:data_reduction:NIRCam}
We used publicly available JWST/NIRCam imaging mosaics of the A2744 field, from the UNCOVER DR3 \citep{Bezanson2024_UNCOVER}, retrieved from the DAWN JWST Archive (DJA\footnote{\url{https://dawn-cph.github.io/dja/index.html}}). The DJA imaging products are homogeneously reduced using the \texttt{grizli} software package \citep{Brammer2023_grizli} and the general NIRCam data reduction procedure is described in \citet{Valentino2023_grizli}. 
Briefly, the DJA reductions are based on the JWST calibrated imaging products and include additional processing to improve the quality of the NIRCam imaging, including corrections to the photometric calibration and detector-dependent variations, masking and mitigation of detector artifacts and cosmic rays, background processing, astrometric alignment, and combination of the individual exposures into drizzled mosaics.
We use the publicly released version v7.2 mosaics for the analysis presented here, with the SW and LW filters reduced at scales of 0.02 and 0.04 arcsec/pixel, respectively.
We successively produce background-subtracted cutouts in the region of our targets, subtracting the diffuse emission from the galaxy outskirts and from the intracluster light of A2744 as a spatially moving median over a box of $2\times2$ arcsec$^2$. We also make use of PSF-matched images, either to F200W (when considering only the SW filters) or to the F444W; the model PSFs are retrieved using the official STPSF tool\footnote{\url{https://stpsf.readthedocs.io/en/latest/index.html}}, while the matching is done using the PSF and convolution tools from the \texttt{photutils} and \texttt{scipy} \texttt{python} libraries. As an example of the resulting data, the stacking of the PSF-matched (to F200W) and background subtracted observations in the SW filters is shown in the top-right panel of Fig.~\ref{fig:fov}.

\section{Analysis and Results}\label{sec:analysis_results}

\subsection{Lens model}\label{sec:lens_model}
The strong lensing model of A2744 ($\rm z=0.3072$) used in this work is described in the companion paper \citet{RojasRuiz2026_arxiv} and here we report only its key information. The model was built with the software \texttt{Gravity.jl} as an update of the parametric model presented by  \citet{Bergamini2023}. The model is constrained by the observed positions of 149 multiple images (121 of which spectroscopically confirmed), associated with 50 background sources and spanning the redshift range $\rm 1.0258 \leq z \leq 9.756$. The observed positions and redshifts of LAP2a and LAP2b are included among the constraints. 
The best-fit model yields a root-mean-square separation $\rm \Delta_{rms}=0''.36$ between the observed and model-predicted multiple-image positions over the entire sample. The observed positions of LAP2a and LAP2b are model-predicted with a precision of $0''.21$ and $0''.30$, respectively. 
Both counter-images are expected to be extremely stretched along one direction, with predicted total (and tangential) magnifications $\rm \mu_{tot}~(\mu_{tan})=51.3^{+4.6}_{-3.6}$ $(21.9^{+3.1}_{-2.4})$ for LAP2a and $\rm \mu_{tot}~(\mu_{tan})=46.9^{+4.1}_{-3.2}$ $(21.3^{+4.2}_{-2.9})$ in the case of LAP2b. Given that LAP2a is the image with the highest magnification and also with the deepest exposure, we use it as a reference for the current study of LAP2. We point out that the study of LAP2b is further complicated by the presence of a brighter galaxy next to it. The analysis of LAP2b is summarized in Appendix~\ref{sec:app:lap2b}.

\begin{table}[]
    \centering
    \caption{Main photometric and spectroscopic properties of LAP2 and of its core region.} 
    \renewcommand{\arraystretch}{1.35}
    \begin{tabular}{l|rr}
    \ & LAP2a & $\rm LAP2a_{core}$ \\
    \hline
    RA & 0:14:21.933 & $''$ \\
    Dec & $-$30:23:54.960 & $''$ \\
    $\rm \mu_{tot}$ & $51.3^{+4.6}_{-3.6}$ & $''$ \\
    $\rm \mu_{tan}$ & $21.9^{+3.1}_{-2.4}$ & $''$ \\
    
    \hline
    \ & \multicolumn{2}{c}{Photometry} \\
    $\rm R_{eff}~[pc]$ & $17.8^{+11.0}_{-8.9}$ & $7.3^{+5.4}_{-3.4}$ \\
    $\rm m_{UV}~[AB]$ & $29.2^{+0.1}_{-0.2}$ & $29.8^{+0.2}_{-0.1}$ \\
    $\rm M_{UV}~[AB]$ & $-12.7^{+0.2}_{-0.2}$ & $-12.1^{+0.2}_{-0.1}$ \\
    $\rm \beta_{UV}$ & $-2.51^{+0.55}_{-1.26}$ & $-3.15^{+0.67}_{-0.69}$ \\
    \hline
    \ & \multicolumn{2}{c}{Spectroscopy} \\
    $\rm z_{sys,H\alpha}$ & $\rm 4.199\pm0.001$ & $4.200\pm0.001$ \\
    $\rm \mu F_{H\alpha}~[10^{-19}~erg~s^{-1}~cm^{-2}]$ & $3.9\pm0.4$ & $1.2\pm0.1$ \\
    EW$\rm _{rest}$($\rm H\alpha$) [\AA] & $880^{+180}_{-100}$ & $570^{+130}_{-80}$ \\
    $\rm log(\xi_{ion}/[erg^{-1}~Hz])$ & $25.3\pm0.1$ & $25.0\pm0.1$ \\
    $\rm \mu F_{H\beta}~[10^{-19}~erg~s^{-1}~cm^{-2}]$ & $1.5\pm0.3$ & $0.5\pm0.1$ \\
    $\rm \mu F_{[OIII]}~[10^{-19}~erg~s^{-1}~cm^{-2}]$ & $\leq0.9~(3\sigma)$ & $0.6\pm0.2$ \\
    $\rm R3$ & $\leq0.61~(3\sigma)$ & $1.18\pm0.45$ \\
    $\rm 12+log(O/H)$ & $\leq6.39~(3\sigma)$ & $6.62\pm0.14$ \\
    $\rm Z~[Z_\odot,\%]$ & $\leq0.5~(3\sigma)$ & $0.9\pm0.3$ \\
    EW$\rm _{rest}$($\rm He II~\lambda1640$) [\AA] & $<39~(3\sigma)$ & $-$ \\
    \hline
    \ & \multicolumn{2}{c}{SED fitting} \\
    $\rm age_{mw}~[Myr]$ & $7^{+11}_{-0}$ & $13^{+7}_{-4}$ \\
    $\rm log(M_\star/[M_\odot])$ & $4.9^{+0.2}_{-0.1}$ & $5.0^{+0.1}_{-0.2}$ \\
    $\rm log(\Sigma_{M_\star}/[M_\odot~pc^{-2}])$ & $1.6^{+0.5}_{-0.6}$ & $2.5^{+0.7}_{-0.7}$ \\
    $\rm A_V~[mag]$ & $0.06^{+0.07}_{-0.04}$ & $0.00^{+0.16}_{-0.00}$ \\
    $\rm log(SFR_{10}/[M_\odot~yr^{-1}])$ & $-2.00^{+0.03}_{-0.28}$  & $-2.59^{+0.47}_{-0.12}$ \\
    $\rm log(\Sigma_{SFR}/[M_\odot~yr^{-1}~kpc^{-2}])$ & $0.7^{+0.4}_{-0.6}$ & $0.9^{+0.8}_{-0.7}$ \\
    %$\rm sSFR$ & & \\
%    \hline
    \hline
    \hline
    \end{tabular}
    %\tablefoot{}
    \label{tab:properties}
\end{table}

\subsection{Morphology and photometry}\label{sec:photometry}
Imaging in the rest-UV reveals that LAP2 is characterized by a compact bright region surrounded by fainter emission (Fig.~\ref{fig:fov}). We fit a combined 2-Gaussians profile to the image obtained by the stack of all SW filters (excluding F070W\footnote{The F070W filter contains the \lya\ emission, which can be geometrically displaced from the stellar continuum.}) PSF-matched to F200W; the fit returns an effective radius of $0.057$ arcsec, corresponding to $\rm \mu R_{eff}=391^{+237}_{-186}~pc$. Considering that the source is unresolved in the radial direction,  assuming a symmetrical source (in the source plane), we use the tangential magnification (as discussed in numerous previous similar studies, see e.g., \citealt{Claeyssens2025}) to derive an intrinsic radius $\rm R_{eff}=17.8~pc$.
If we instead fit the source assuming a single Sersic profile, reaching a similarly good fit (for $\rm n_{Sersic}\sim4$), we would derive a slightly larger radius, $\rm R_{eff,1S}=24~pc$. 
The morphology of the core region alone appears very compact in the current data, hinting at a very small intrinsic radius, $\rm R_{eff,core}=7.3~pc$.
%\footnote{Assuming, for the unresolved case that $\rm R_{eff}\equiv 0.5\times FWHM_{PSF}/\mu_{tang}$, see e.g. \citet{Claeyssens2025}.}.

We perform aperture photometry on PSF-matched data to the reddest filter, F444W, using the apertures shown in Fig.~\ref{fig:fov}. We apply aperture correction to account for the missed flux outside the apertures ($-0.2$ and $-0.9$ mag for LAP2 and its core, respectively). 
%The correction assumes an unresolved source at the center of the aperture, justified by the central compact region dominating the flux.
%; this is accurate for the aperture covering the compact core, but is a slight underestimation for the larger aperture. \red{Maybe we can compare to estimates coming from the size-phot? (by quick test they have too low signal in some bands to be reliable).}
The apparent UV magnitude (measured in F090W, corresponding to $\sim1700~\AA$) of LAP2a is $\rm m_{UV}=29.2$ ($\rm m_{UV}=29.8$ for the compact core), corresponding to intrinsic magnitude $\rm M_{UV}=-12.7$ ($\rm M_{UV}=-12.1$ for the compact core) when accounting for distance and magnification. This makes LAP2 comparable to the faintest observed galaxies at high-z in GLIMPSE, the lensed field with the deepest images to date \citep{Atek2026_arXiv, Korber2026}.

We evaluate the UV slope using the F090W, F115W and F150W filters, thus encompassing the wavelengths $1500-3200~\AA$; as mentioned above, F070W is excluded because it contains the \lya\ emission. LAP2 has an overall blue slope $\rm \beta_{UV}=-2.5$ that gets steeper at the position of the compact emission, $\rm \beta_{UV,core}=-3.1\pm0.7$, even if with large associated uncertainties. These values are steeper than the average $\rm \beta_{UV}$ observed at $\rm z\gtrsim4$ \citep[e.g.,][]{Topping2024}, even if extreme slopes down to $\rm \beta<-3$ are being routinely observed with JWST \citep[e.g.,][]{Topping2022,Cullen2023,Cullen2024}. We point out that both fainter galaxies and compact UV clumps, with radii similar to LAP2a, have bluer slopes than the average \citep{Bolamperti2023}. 
%\red{Discuss the photometry of the large aperture, with a strange shape?}

\subsection{IFU spectroscopy}\label{sec:spectroscopy}
\begin{figure*}
    \centering
    \includegraphics[width=0.99\linewidth]{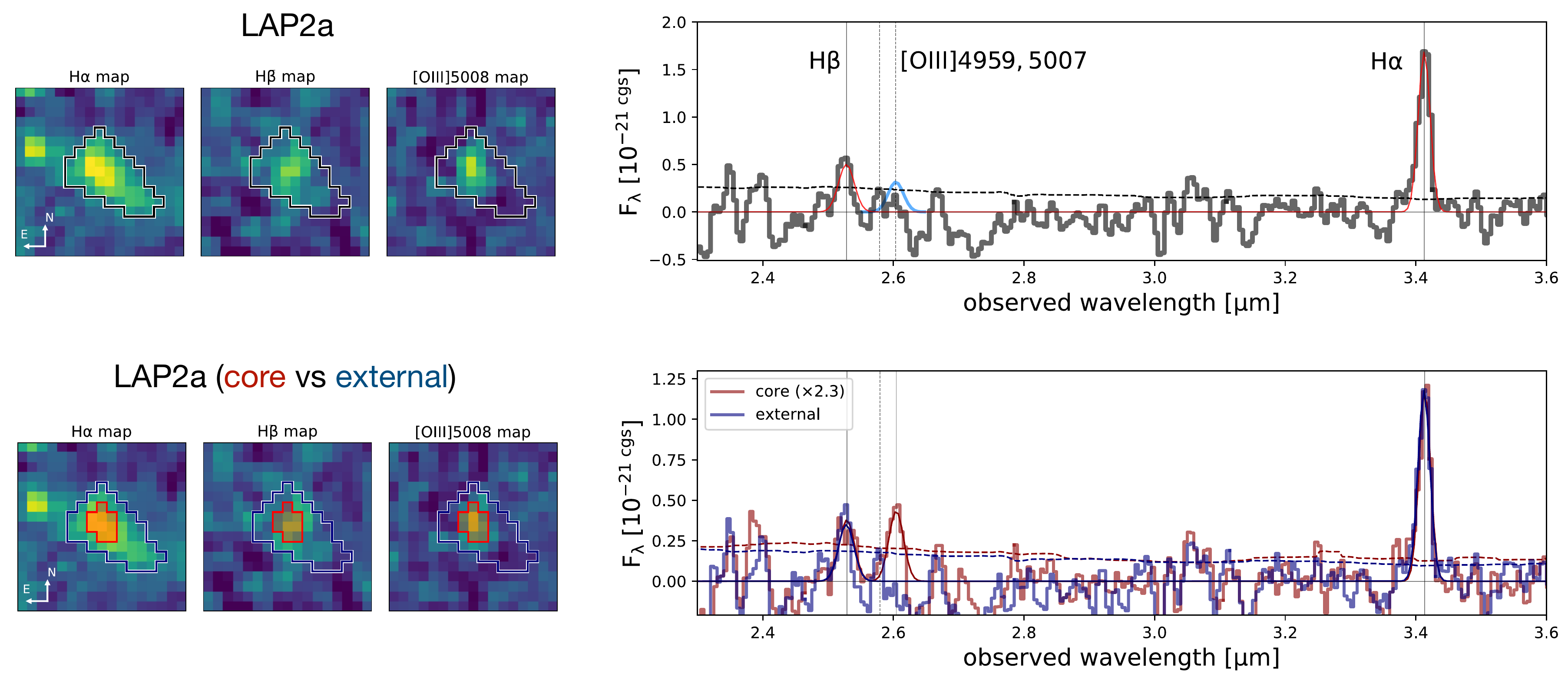}
    \caption{Emission line maps (left panels, with a scale of 0.05 arcsec/pixel) and extracted spectra between the \hb\ and the \ha\ wavelengths (right panels). The top panels show an extraction covering the full LAP2a emission; the spectrum, uncertainty spectrum and best-fit spectrum are plotted as solid wide gray, dashed black and solid red lines, respectively. A mock \oiiiv\ line at $\rm 3\sigma$ level is added as a blue line, for comparison.
    Bottom panels show the comparison between spectra extracted from the `core' region (in red) and from the total mask, excluding the core (`external' region, in blue). The `core' spectrum was rescaled to match the \ha\ flux of the external region; while the spectra at the \ha\ and \hb\ wavelengths almost overlap, only the core spectrum shows a \oiiiv\ detection.}
    \label{fig:spec_main}
\end{figure*}

The $\rm H\alpha$ emission of LAP2 is clearly detected in the IFU observations, at a redshift $\rm z_{H\alpha}= 4.199 \pm 0.001$. From the two-dimensional \ha\ map, the emission shows a bright compact component surrounded by a fainter one (see Fig.~\ref{fig:spec_main}, left panel), similarly to what observed in the stellar continuum with NIRCam. 
The \ha\ emission map shows also an additional faint component, towards the north-east side of LAP2a, without any clear counterpart in the stellar continuum; its line profile is slightly irregular and the emission is therefore excluded from the main analysis. a more detailed description of this extra component is given in Section~\ref{sec:extra_ha}.

The total observed \ha\ flux of LAP2a, $\rm \mu F_{H\alpha}=(3.9\pm0.4)\cdot10^{-19}~erg~cm^{-2}~s^{-1}$, translates into a rest-frame equivalent width $\rm EW(H\alpha)=880^{+180}_{-100}~\AA$. The line appears spectroscopically unresolved in our prism observations\footnote{The PRISM resolution at the observed wavelength of \ha\ is $\rm FWHM\leq0.026~\mu m$, corresponding to $\rm \sigma_v\leq2300~km~s^{-1}$.}. When compared to the UV flux observed with NIRCam (in F090W, rest-frame $\sim1700~\AA$), we derive a ionizing radiation production efficiency $\rm log(\xi_{ion})=25.3\pm0.1$ (assuming zero escape fraction and no attenuation). If instead the UV continuum were estimated at rest-frame $1500~\AA$ using the best-fit SED model, its value would be highly dependent on the fit assumptions (Section~\ref{sec:SED}); the resulting ionization efficiency parameter would vary in range $\rm log(\xi_{ion})=25.2-25.5$.

An emission line at the expected wavelength of \hb\ is observed at low signal-to-noise ratio. The line appears slightly broader than the spectral resolution, possibly due to noise; we therefore fit the line fixing both the wavelength (according to the redshift inferred from \ha) and the line width (fixing it to the instrumental resolution). The best-fit returns a flux $\rm \mu F_{H\beta}=(1.5\pm0.3)\cdot10^{-19}~erg~cm^{-2}~s^{-1}$, giving a Balmer decrement $\rm H\alpha/H\beta=2.5\pm0.5$, consistent with no extinction. 
Finally, no \oiii\ line is observed above the detection threshold. We derive a $\rm 3\sigma$ limit on the flux of the \oiiiv\ line, as (three times) the root sum square of the error spectrum in 6 channels around the expected wavelength\footnote{The line resolution at the expected \oiiiv\ wavelength is $\rm FWHM=5.7\times\Delta x$, where $\rm \Delta x=0.005~\mu m$ is the width of a spectral channel.}, finding $\rm \mu F_{[OIII]}<9.3\cdot 10^{-20}~erg~cm^{-2}~s^{-1}$, approximately a factor $0.6$ of the \hb\ flux. The expected line from this limit is shown with a blue color in the spectrum of Fig.~\ref{fig:spec_main} (top panel); according to the strong-line calibration by \citet{Sanders2024} (extrapolated to extremely low R3 values), the $\rm 3\sigma$ upper limit on the R3 ratio ($\rm R3\equiv [OIII]5007/H\beta\leq0.61$) translates into a metallicity $\rm 12+log(O/H)\leq6.4$ (or $\rm \lesssim0.5\%~Z_\odot$). 
%\red{Adding an estimate from Ha assuming a best-fit extinction?}

While LAP2a appears overall oxygen-deficient, the IFU map at the expected wavelength of the \oiiiv\ emission shows a very faint signal in correspondence to the compact component of the stellar (and \ha) emission (Fig.~\ref{fig:spec_main}). We thus define a new smaller mask covering the \oiii\ signal (Fig.~\ref{fig:spec_main}, bottom panel), co-spatial to the position of the stellar continuum core analyzed in Section~\ref{sec:photometry}. By extracting the spectrum in this ``core mask'' we recover $\sim30\%$ of the \ha\ and \hb\ fluxes (with respect to the main mask) but we also detect a \oiiiv\ line with flux $\rm \mu F_{[OIII]}=(5.6\pm1.5)\cdot 10^{-20}~erg~cm^{-2}~s^{-1}$. 
The detection of oxygen emission in the core of LAP2a points towards a larger metallicity, which we estimate again from the \citet{Sanders2024} strong-lines calibration; in this case the R3 ratio, $\rm R3=1.2\pm0.4$ converts into $\rm 12+log(O/H)=6.6\pm0.1$. Given the low signal of the \hb\ emission, we can alternatively estimate the R3 index starting from the \ha\ emission (assuming the case-B recombination and no extinction ratio $\rm H\alpha/H\beta=2.86$), deriving $\rm 12+log(O/H)=6.7\pm0.1$, confirming that the metallicity of the core of LAP2a is close to $\rm 1\%~Z_\odot$. When considering a mask that excludes the core region, \oiiiv\ remains undetected (see Fig.~\ref{fig:spec_main}, bottom panels) and the $\rm 3\sigma$ upper limit on the R3 index, $<0.65$, remains consistent with the value found for the total mask (Fig.~\ref{fig:spec_main}, top panels).
The spatial variation in R3, and possibly oxygen abundance, is further discussed in Section~\ref{sec:discussion:spatial_metallicity}.

As previously mentioned, while the counterimage \textit{b} (i.e. LAP2b) is covered by the IFU observations, it provides much looser constraints on the metallicity measurement of the source, mainly because of its low observed flux (due to lower magnification). The closeness of LAP2b to the edge of the field-of-view further implies a more noisy spectrum. More details on the spectral analysis of LAP2b are given in Appendix~\ref{sec:app:lap2b}.

Finally, He II emission is usually invoked as a constraining diagnostic for Pop III stars, as a probe of very hard ionizing spectrum from massive metal-free stars. The $\rm 3\sigma$ flux limit we derive for He II 1640, corresponding to equivalent width $\rm EW_{rest}<39~\AA$, does not provide strong constraints, as Pop III selection criteria in the literature typically require $\rm EW\gtrsim10-20~\AA$ \citep[e.g.][see also discussion in Section~\ref{sec:discussion:metalpoor}]{schaerer2003,Katz2023,Katz2024}.

\subsection{SED fitting}\label{sec:SED}
We fit the broadband SED of LAP2a obtained in Section~\ref{sec:photometry} with \texttt{Bagpipes} \citep[v1.3.6,][]{Carnall2018,Carnall2019} using the BPASS \citep{BPASS} stellar models and assuming a \citet{Kroupa2001} IMF. LAP2 is observationally very faint, with large photometric uncertainties; as a consequence, many of the fit parameters may remain unconstrained due to degeneracies. We use a \citet{Calzetti2000} attenuation law, and assume a parametric SFH (delayed-$\tau$). We fix the ionization parameter to $\rm log(U)=-2$, the metallicity to $\rm 1\%~Z_\odot$ and the escape fraction to $\rm f_{esc}=0$. The only remaining free parameters are the age and $\tau$ parameters describing the SFH, the stellar mass formed and the extinction (parametrized by $\rm A_V$). 
The best fit of the broadband SED returns a relatively old  system (mass-weighted age $\rm age_{mw}=27~Myr$) characterized by a short burst of star formation ($\rm \tau\approx1~Myr$, though poorly constrained, extending to $\rm \tau\approx50~Myr$ within $1\sigma$). This SFH would imply null SFR within the last few Myr, inconsistent with the line emission observed in the IFU spectrum (Section~\ref{sec:spectroscopy}). 
%\red{It could be worth testing a fit with non-parametric SFH, even if not expecting major differences. TESTED, no major differences, not included} 
%\red{Is it worth to add a figure of the SED fitting? UGLY photometric SED, maybe not worth it.}
We thus include in the SED fitting process the part of the spectrum containing the \ha\ and \hb\ emission lines, in order to better constrain the recent SFH of the system. In this case, we leave the metallicity free to let the model adjust to the observed line ratios. The resulting best-fit age is younger than previously derived ($\rm age_{mw}=7~Myr$) with a SFH still characterized by a single burst scenario, implying an average $\rm \mu SFR=0.5~M_\odot/yr$ over the last $\rm 10~Myr$. Similarly, a $\rm 0.4~M_\odot/yr$ value would be found by converting the observed \ha\ flux into SFR according to the \citet{Kennicutt2012} calibration. 
The fit confirms that LAP2 is a low-mass system, with a magnified stellar mass $\rm log(\mu M_\star/M_\odot)=6.6$, corresponding to an intrinsic mass $\rm log(M_\star/M_\odot)=4.9$; as previously noticed in case of the compact size, in terms of mass LAP2 is comparable to massive star clusters, observed both locally \citep[e.g.,][for a review]{krumholz2019} and at high-z \citep[e.g.,][among the most recent]{Adamo2025,Nakane2025_arXiv,Claeyssens2026_arXiv}.
The system appears to have only low extinction ($\rm A_V\lesssim0.1$ mag), consistently with the Balmer decrement measured from the spectrum.
Finally, we point out that the best-fit model under-predicts the flux in the bluest bands used (F090W and F115W). We test a fit with $\rm f_{esc}$ as an additional free parameter; the best-fit model ($\rm f_{esc}=77\%$) is characterized by an extremely young age ($\rm age_{mw}=2~Myr$), but its mass remains comparable to the previous case. 

We perform SED fitting to the spectrum and the photometry also in the case of the core region of LAP2a, assuming that the small mask and aperture shown in Figs.~\ref{fig:fov} and~\ref{fig:spec_main} are tracing emission from the same physical region (despite the difference in spatial resolution). In this case, the recovered age in the $\rm f_{esc}=0$ case ($\rm age_{mw}=13~Myr$) remains consistent when the escape fraction is left free (best fit $\rm f_{esc}=60\%$, $\rm age_{mw}=10~Myr$). The mass remains consistent with that derived from the larger aperture; this is reasonable, considering that the small aperture contains more than half of the total flux. 

Despite the small intrinsic size of LAP2a, its low mass and SFR correspond to low surface densities $\rm log(\Sigma_M/[M_\odot~pc^{-2}])=1.6^{+0.5}_{-0.6}$ and $\rm log(\Sigma_{SFR}/[M_\odot~yr^{-1}~kpc^{-2}])=0.7^{+0.4}_{-0.6}$; those values are smaller than the average surface densities observed in compact star-forming regions at $\rm z\sim4$ \citep[see e.g.,][]{Messa2024,Claeyssens2025}. In order to obtain values closer to those of its high-z counterparts, one needs to consider the core of LAP2a, reaching higher surface densities, $\rm log(\Sigma_M/[M_\odot~pc^{-2}])=2.5^{+0.7}_{-0.7}$ and $\rm log(\Sigma_{SFR}/[M_\odot~yr^{-1}~kpc^{-2}])=0.9^{+0.8}_{-0.7}$.

\subsection{The ``extra'' $\rm H\alpha$ emission}\label{sec:extra_ha}
\begin{figure*}
    \centering
    \includegraphics[width=0.39\linewidth]{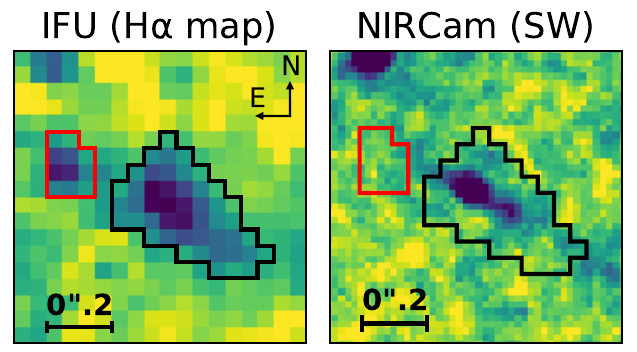}
    \includegraphics[width=0.60\linewidth]{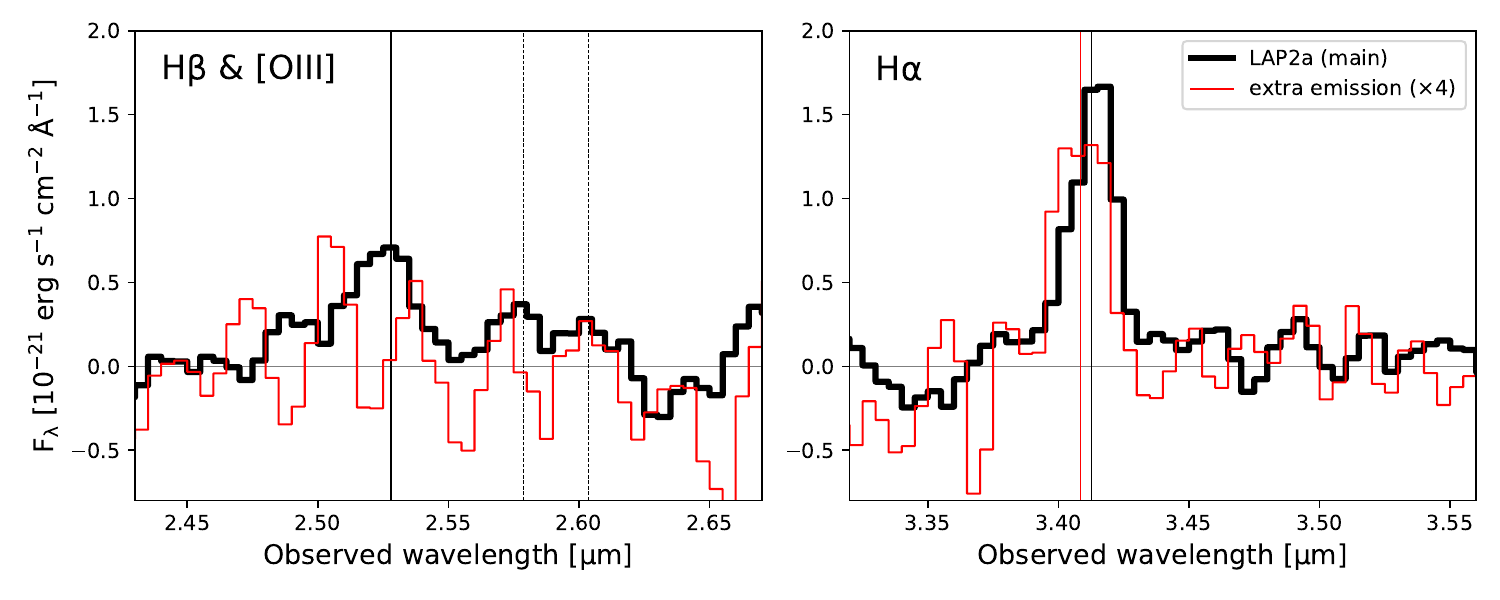}
    \caption{\textit{Left panels:} comparison between the emission line map (from NIRSpec/IFU) and the stellar continuum map (from NIRCam), highlighting the spatial location of the `extra' \ha\ emission. \textit{Right panels:} spectral comparison, at the wavelengths of \ha\ and \hb, between the main (in black) and the `extra' (in red) components.}
    \label{fig:extra_ha_emission}
\end{figure*}

As previously noticed in Section~\ref{sec:spectroscopy}, the \ha\ map shows a further possible emission north-east of the main LAP2a system, as shown by the red contour in Fig.~\ref{fig:extra_ha_emission}. In the figure we overlay its position on the stellar continuum map and we show its spectrum compared to the one extracted from the main aperture. 
Two details question the authenticity of this emission: 1) the emission is spatially detached from the observed stellar continuum, by $\gtrsim0.1$ arcsec, which correspond to $\sim30$ pc at the redshift of the source, assuming de-lensing via the tangential magnification; 2) the line profile appears flatter around the peak (with a small plateau), and slightly blue-shifted, by $\rm 44\pm16~\AA$ (corresponding to a redshift difference of $\rm 0.007\pm0.002$, or a velocity difference of $\rm 390\pm140~km~s^{-1}$) than the reference \ha\ from the main aperture. This possible emission is detected with SNR$\gtrsim5$; no \hb\ is detected within the same mask, although its expected flux (based on \ha) would be comparable to the noise level. If the flux from this mask, covering the extra emission, were considered in the total spectrum, the Balmer decrement would increase to $\rm H\alpha/H\beta=2.97$, still consistent with no (or very low) attenuation, while the R3 index would remain almost unchanged. 

%--------------------------------------------------------------------

\section{Discussion} \label{sec:discussion}

\subsection{How metal-poor is LAP2?}\label{sec:discussion:metalpoor}
The non-detection of [\ion{O}{iii}]$\lambda5007$ in the integrated spectrum confirms LAP2 as an oxygen-deficient system, as first suggested by the MSA
spectrum of LAP2b \citep{Vanzella2026b}. Translating a low R3 into a low oxygen abundance, however, requires two caveats to be addressed. First, at fixed O/H, R3 decreases with decreasing ionization parameter \citep[e.g.,][]{Kewley2019,nakajima2022}, so a weak [\ion{O}{iii}] line could in principle reflect a soft or dilute radiation field rather than a low metal content. This scenario is disfavored for LAP2: the large EW(H$\alpha)\simeq880$ \AA \, and $\log\xi_{\rm ion}\simeq25.3$ indicate the presence of a young stellar population (few Myr old), excluding  a low-ionization nebula \citep[e.g.,][]{Atek2022}. Second, strong-line calibrations are not constrained by direct-$\rm T_e$ data below $12+\log({\rm O/H})\approx7$, so our values rely on an blind extrapolation (of the \citealp{Sanders2024} calibration); the corresponding uncertainty, at least
$\sim0.1$ dex from the scatter in R3 at fixed metallicity alone, is likely larger in practice. With these caveats, the integrated limit $12+\log({\rm O/H})\lesssim6.4$ places LAP2 among the most metal-poor systems known at any redshift (see Section~\ref{sec:discussion:literature}), though still orders of magnitude above the critical metallicity expected for Pop III star formation ($\lesssim10^{-3.5}\,Z_\odot$, e.g., \citealp{Bromm2004}). Its emission is thus consistent with an extremely metal-poor population, possibly mixed with a pristine component \citep[e.g.,][]{Venditti2023}.
Consistently, the upper limit on \ion{He}{ii} (EW(1640) $<39$ \AA) does not exclude a Pop III contribution, ruling out only the most extreme metal-free bursts (e.g. for a zero-age, top-heavy IMFs, the EW can reach up to $\sim50-100~\AA$, \citealp{schaerer2002,schaerer2003}). 

\subsection{Spatial variations and the enrichment budget}\label{sec:discussion:spatial_metallicity}
A key advantage of the IFU observations over the previous NIRSpec/MSA spectrum of LAP2b \citep{Vanzella2026b} is that they capture the full emission of the system and resolve it spatially. A single MSA shutter samples only part of a strongly lensed, extended source, and mixes the emission of all components falling within it; an integrated line ratio can therefore hide significant variations on small scales, which are expected in systems where pristine and already enriched gas coexist. Such variations have been found on small scales in other high-redshift sources \citep[e.g.,][]{Carniani2026_arXiv}. Low-metallicity regions have also been found adjacent to brighter, more chemically evolved systems \citep[e.g.,][]{Maiolino2024,Ubler2026,Vanzella2024_t2c}, as also predicted by theoretical models \citep[e.g.,][]{Venditti2023,Venditti2026}. Spatially resolved spectroscopy is therefore essential both to locate the most pristine regions and to avoid diluting them with emission from enriched neighbors.

The IFU data of LAP2 show that this is indeed the case: the picture is not uniform. Faint [O\,{\sc iii}] emission is detected only in the compact core, which corresponds to $12+\log({\rm O/H})=6.6\pm0.1$ there, compared with $\leq6.4$  integrated over the entire galaxy. Because the core is more compact and hosts a higher stellar surface density, it is also expected to have a higher ionization parameter; at these metallicities this alone could raise R3 without any change in O/H \citep[e.g.,][]{Kewley2019}. A modest metallicity gradient and an ionization gradient are therefore both viable, and not mutually exclusive.

To further complicate the picture, recent studies have shown that high electron densities can suppress collisionally excited forbidden lines through collisional de-excitation, potentially leading to an underestimate of the gas-phase oxygen abundance when standard low-density assumptions are adopted. For [\ion{O}{iii}]$\lambda5007$, collisional suppression becomes increasingly important only for $\rm n_e\gtrsim10^{6}\ {\rm cm^{-3}}$ \citep[e.g.,][]{Martinez2025}.
While we do not have a direct probe of the density in LAP2, we point out that its surface density (in both stellar mass and SFR) is not particularly high, as it would have been reasonable to expect in the case of high $\rm n_e$ \citep[e.g.,][]{Isobe2023,Citro2024,Laseter2024,Boyett2024,Topping2024}. 
If the difference in R3 traces a genuine difference in oxygen abundance, it is straightforward to explain by local self-enrichment. Given that the H$\alpha$ luminosity of the core implies an ionized gas mass of $\sim10^4\,(n_e/100\,{\rm cm^{-3}})^{-1}\,{\rm M_\odot}$\footnote{Assuming case B and $\rm T_e=10^4~K$ \citep{osterbrock2006}.}, raising the gas from $12+\log({\rm O/H})\lesssim6.4$ to $6.6$ requires only $\sim0.2\,{\rm M_\odot}$ of newly synthesized oxygen. Assuming standard IMF-integrated oxygen yield from SNe ($\sim5\times10^{-3}$--$10^{-2}$, e.g. \citealp{Nomoto2006}) this corresponds to the ejecta of $\sim10^2\,{\rm M_\odot}$ of stars, i.e. a negligible fraction ($\sim0.1\%$) of the $\sim10^5\,{\rm M_\odot}$ formed in the core.
LAP2 is a very young system (mass-weighted age $\lesssim10$ Myr) thus the main uncertainty may be the timescale of metal mixing, as in local (dwarf) galaxies most supernova ejecta are thought to
remain in a hot phase for $\sim10^8$ yr before mixing into the warm ISM \citep[e.g.,][]{TenorioTagle1996,Kobulnicky1997,Hirai2017}.
However, since only $\sim0.1\%$ of the newly synthesized oxygen needs to be incorporated into the warm ionized gas to explain the observed difference, even an inefficient and partial mixing within the age of the core would suffice. A more quantitative assessment of the cooling and mixing of the ejecta in such a compact system is beyond the scope of this work. 

\subsection{Ionizing conditions and feedback}
Several independent observables point to a hard and partly escaping ionizing radiation field. The UV slope ($\beta\simeq-2.5$, and possibly steeper, down to $\beta\lesssim-3$, in the core) approaches or exceeds the values expected for stellar populations, including nebular continuum. The SED models with $f_{\rm esc}=0$ under-predict the bluest bands, and solutions with free $f_{\rm esc}$ prefer values $\simeq60-80\%$. If taken at face value, such escape fractions would raise the intrinsic $\xi_{\rm ion}$ to $\log\xi_{\rm ion}\sim25.7$--$25.9$, approaching values predicted for extremely metal-poor and Pop III populations \citep[e.g.,][]{schaerer2003,Raiter2010}. 
The \ha-only "extra" component north-east of LAP2a (Section~\ref{sec:extra_ha}), if confirmed, could be a natural candidate for the reprocessing of the core radiation: a gas cloud a few tens of pc from the core, ionized by escaping radiation, with no associated stellar continuum. An alternative explanation is a faint, very young stellar clump below the NIRCam detection limit, the scenario envisaged for late Pop III star formation in pristine pockets around star-forming systems \citep[e.g.,][]{Venditti2023}. 
\begin{figure*}
    \centering
    \includegraphics[width=0.98\textwidth]{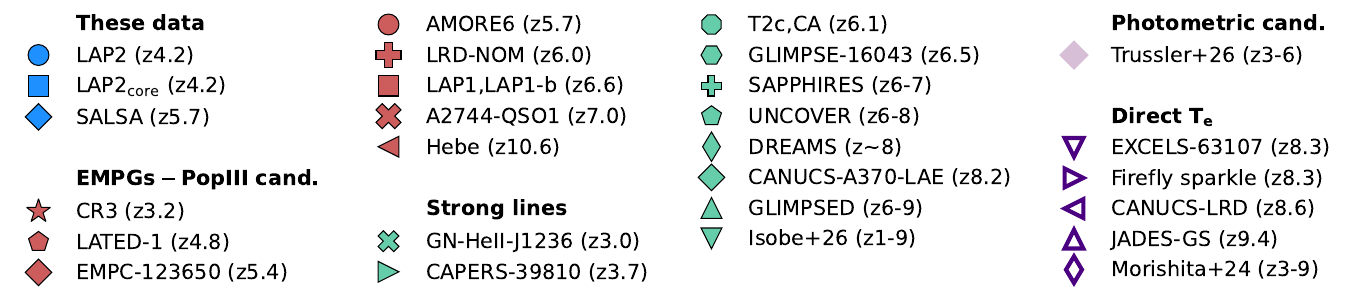}
    \includegraphics[width=0.99\columnwidth]{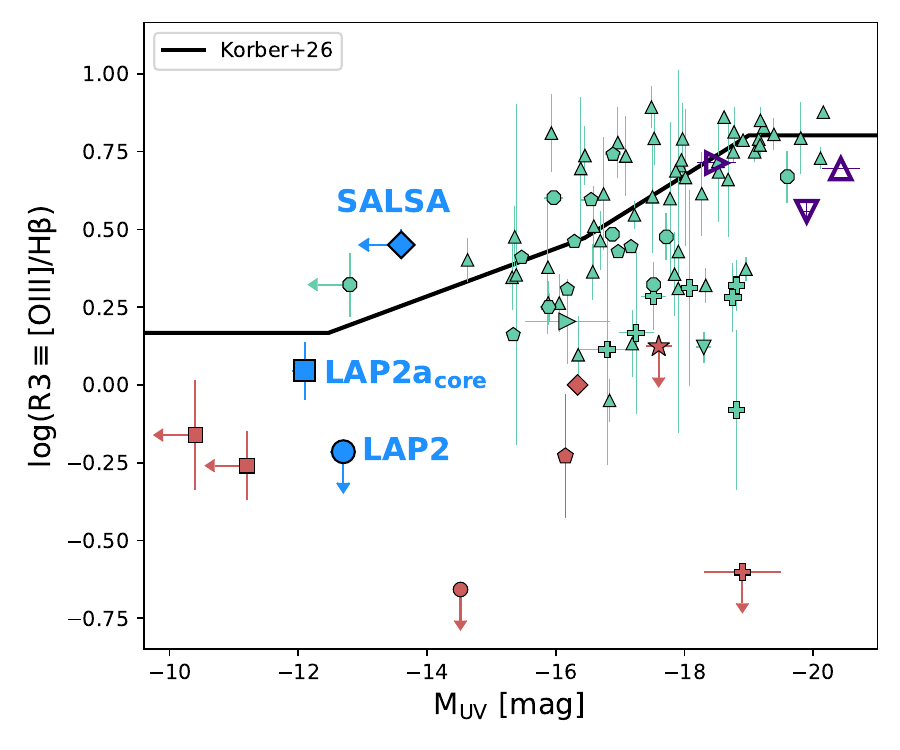}
    \includegraphics[width=0.99\columnwidth]{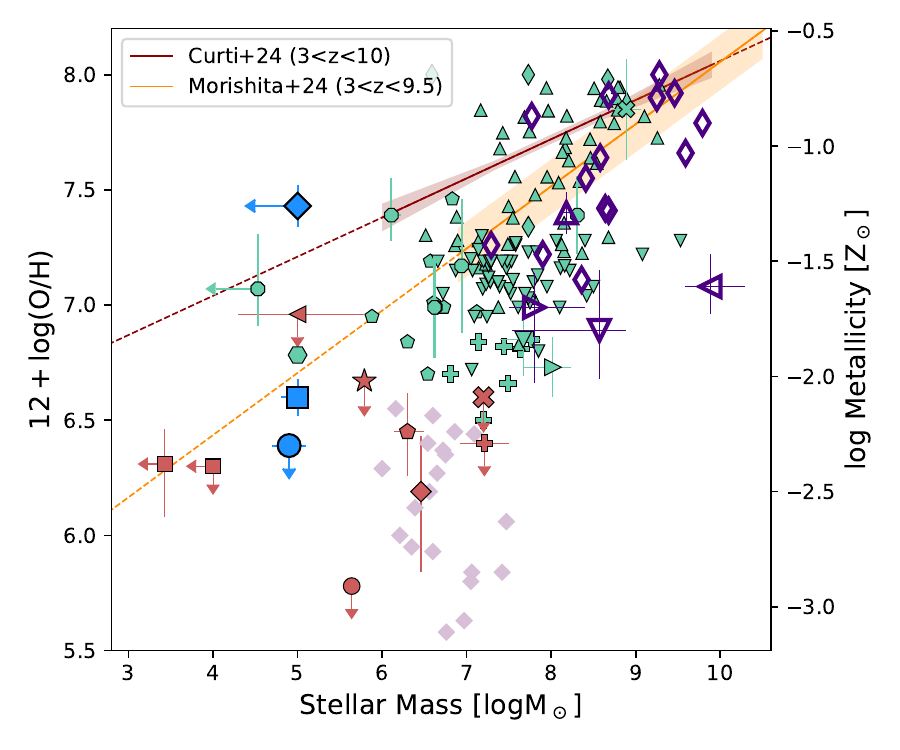}
    \caption{Rest-frame UV luminosity vs R3 index ({\it left panel}) and stellar mass vs oxygen abundance ({\it right panel}) of LAP2 and the SALSA arc at $z=5.66$ compared to literature systems. The latter were chosen among known high-redshift ($z>3$) galaxies, focusing on low-luminosity ($\rm M_{UV}\gtrsim-19$) and low-metallicity systems. The sample is color-coded according to the methodology used to infer the metallicity (strong lines vs direct $\rm T_e$ methods, vs photometric candidates). It is further divided to visually separate the lowest metallicity sources, possibly consistent with being close to pristine (e.g., $\rm Z\lesssim1\%~Z_\odot$ or $\rm R3\lesssim1$). We point out that the ``strong line'' sample is inhomogeneous, as different empirical calibrations are used: \citet{Sanders2024} for, e.g., LAP2; \citet{Isobe2026_arXiv} for LATED-1;
    Cloudy \citep{Ferland2017} and BPASS \citep{Stanway_Eldridge2023} photo-ionization models  for AMORE6; \citet{nakajima2022} for, e.g., GLIMPSE-D; Bagpipes \citep{carnall19} SED fitting for, e.g., EMPC-123650. Masses also may be inhomogeneous, as derived under different assumptions; we however point out that most of the sample assumes either a \citet{Kroupa2001} or a \citet{Chabrier_2003} IMF.
    The references for the literature data are: \citet{RojasRuiz2026_arxiv} (SALSA); \citet{Cai2025} (CR3); \citet{Li2026_arXiv} (LATED-1); \citet{Trussler2026} (EMPC-123650); \citet{Morishita2025} (AMORE6); \citet{Caputi2026} (LRD-NOM); \citet{Vanzella2023_lap1,Nakajima2025} (LAP1 and LAP1B); \citet{Maiolino2026} (A2744-QSO1); \citet{Ubler2026} (Hebe); \citet{Mondal2025} (GN-HeII-J1236); \citet{Cai&Yu2026} (CAPERS-39810); \citet{Vanzella2024_t2c,Bolamperti2026} (CA and T2c); \citet{Fujimoto2025_arxiv} (GLIMPSE-16043); \citet{Hsiao2025} (SAPPHIRES); \citet{Chemerynska2024} (UNCOVER); \citet{Nishigaki2025_arXiv} (DREAMS); \citet{Willott2025} (CANUCS-A370-LAE); \citet{Korber2026} (GLIMPSE-D); \citet{Cullen2025} (EXCELS-63107); \citet{Mowla2024} (Firefly Sparkle); \citet{Tripodi2024} (CANUCS-LRD); \citet{Curti2025} (JADES-GS). Upper limits from the literature have been homogenized to the $3\sigma$ level whenever the original significance was reported; for pseudo-LRD-NOM the significance of the limit is not specified \citep{Caputi2026}.}
    \label{fig:MUV_R3_MZ}
\end{figure*}

\subsection{Comparison to the literature} \label{sec:discussion:literature}
In Fig.~\ref{fig:MUV_R3_MZ}, we compare LAP2 with other high-redshift ($z>3$) galaxies in the R3--$\rm M_{UV}$ plane (left) and in the mass--metallicity plane (right). Both panels focus on low-luminosity and low-metallicity systems. To allow a consistent comparison, all upper limits are homogenized at the same significance level ($3\sigma$). 

R3 is one of the most widely used strong-line indices to probe the metallicity of high-redshift galaxies, especially at low metallicities, where auroral lines are too faint to be detected within reasonable exposure times. At first order, the R3--$\rm M_{UV}$ distribution therefore traces how chemical enrichment proceeds with decreasing luminosity. Collecting the faintest high-redshift galaxies known to date, \citet{Korber2026} find that R3 decreases towards low luminosities and then flattens into a plateau at $\rm M_{UV}\gtrsim-12.5$. 
LAP2 is one of the few systems with R3 constraints at $\rm M_{UV}>-13$. Its integrated upper limit lies $\gtrsim0.35$~dex below the plateau, and LAP1 \citep{Vanzella2023_lap1,Nakajima2025} sits at a similar offset. In contrast, the core of LAP2 is consistent with the plateau. These systems may hint at a further decrease of R3 at the lowest luminosities. However, the relation in this regime relies on very few objects, and the sample is likely biased towards low R3 values, since most of these sources were targeted precisely because they were candidate metal-poor systems.
The SALSA arc at $z=5.66$, which falls in the same IFU field of view, lies at a similar luminosity but above the plateau; its properties are discussed in a separate work \citep{RojasRuiz2026_arxiv}.
 
The right panel of Fig.~\ref{fig:MUV_R3_MZ} gives a more physical view, showing the most metal-poor, low-mass galaxies currently known at high redshift. For most systems (filled symbols) the metallicity is derived from strong-line calibrations, while empty symbols with thick edges mark direct-$T_e$ measurements. Because different strong-line calibrations are used, and all of them are extrapolated below $12+\log({\rm O/H})\approx7$, systematic differences of $\gtrsim0.3$~dex between samples are expected \citep[e.g.][]{Cai2025}. 
For the \citet{Korber2026} sample, where stellar masses are not provided, we estimate them from $\rm M_{UV}$ assuming a constant UV mass-to-light ratio, calibrated on empirical $\rm M_\star-M_{UV}$ relations at $z\sim6-7$ \citep{Gonzalez2011,Duncan2014,Song2016}, implying large mass uncertainties, $\sim0.5$~dex. We also show the photometric metallicity estimates of \citet{Trussler2026}, which still lack spectroscopic confirmation (except EMPC-123650, highlighted in the figure). They show how photometric searches can increase the number of EMPG candidates in this poorly explored regime. Most of the samples in Fig.~\ref{fig:MUV_R3_MZ} were published in the last 2--3 years, reflecting the growing effort to populate the low-mass, low-metallicity end of the MZR.
 
LAP2 is among the most metal-poor galaxies known to date and one of the few with a stellar mass comparable to that of individual star clusters. Its location relative to the MZR depends on which relation is extrapolated, since none is calibrated at these masses: the lowest-mass galaxies in \citet{Morishita2024} have $\rm M_\star>10^7~M_\odot$, more than 100 times the mass of LAP2. With respect to the extrapolation of the \citet{Morishita2024} MZR ($\rm 3<z<9.5$), the integrated limit of LAP2 lies only $\sim0.15$--$0.2$~dex below the relation, similarly to other low-mass EMPGs such as LAP1, while the core lies on it. The shallower \citet{Curti2024} relation instead places LAP2 $\gtrsim0.5$~dex below. At present, therefore, LAP2 is not clearly an outlier from the extrapolated MZR; the main open question is how steeply the MZR continues below $\rm 10^7~M_\odot$.

No clear redshift trend emerges among the known high-redshift EMPGs. Candidates are found from $z\approx3$ (CR3 at $z=3.19$; \citealp{Cai2025}) and $z\approx4$ (LAP2) up to $z=10.6$ (Hebe; \citealp{Ubler2026}). The current sample, however, is too small and too heterogeneous to test for any evolution. Most low-luminosity (low-mass) systems in Fig.~\ref{fig:MUV_R3_MZ} come from lensing fields, which have been targeted with different instruments and selection criteria, so the sample is not representative of the underlying population. On the other hand, lensing, coupled with deep observations, is necessary to reach the smallest and faintest systems where pristine populations are expected.   

\subsection{Large-scale environment} \label{sec:discussion:environment}
\begin{figure}
    \centering
    \includegraphics[width=\columnwidth]{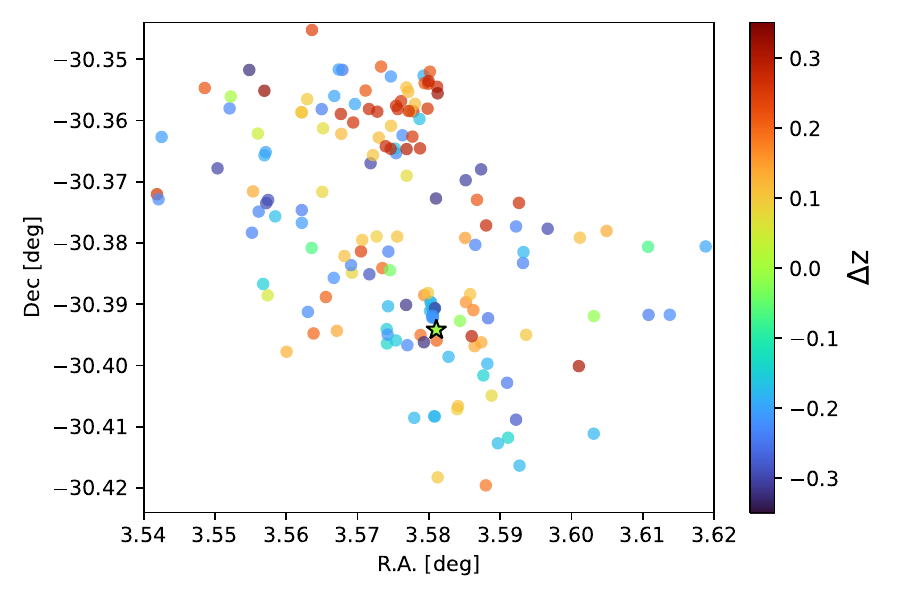}
    \includegraphics[width=\columnwidth]{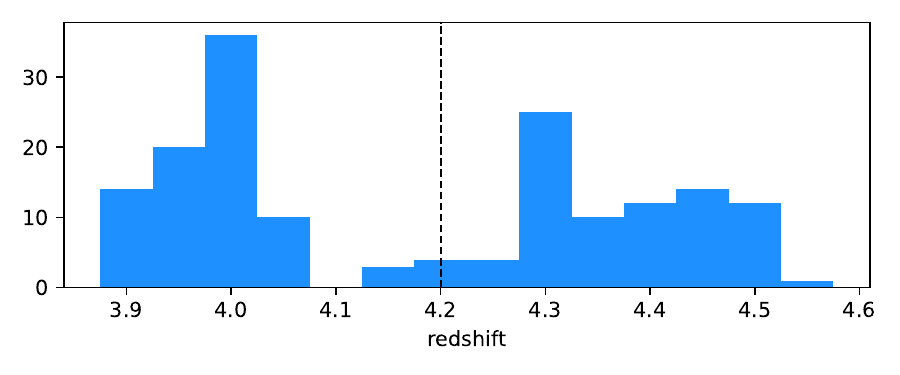}
    \caption{\textit{Top:} (source-plane) projected distribution of galaxies in redshift range $3.85\leq z\leq4.55$, within a region of $\rm \sim2\times2~pMpc^2$ (i.e. covering the main region of the A2744 cluster). See main text for a description of the spectroscopic catalog.
    The galaxies are color-coded according to their redshift. The position of LAP2 is marked with a `star' symbol. \textit{Bottom:} Redshift distribution of the same sources as in the top panel. The redshift of LAP2 is indicated with a dashed line.}
    \label{fig:environment_density}
\end{figure}

To place LAP2 in the context of its large-scale environment, we combine all currently available spectroscopic redshift catalogs of A2744, starting from the catalog of \citet{Bergamini2023}, used to build the lens model, and adding: GLASS-JWST/NIRSPEC \citep{Mascia2024};  JWST/NIRSPEC GO-3073 \citep{Castellano2024}; JWST/NIRCAM WFSS \citep{Naidu2024_arXiv}; DAWN JWST Archive, including \citet{Dottorini2025}; GLASS-JWST/NIRISS WFSS \citep{Watson2025_GLASS}; UNCOVER JWST/NIRSPEC \citep{Price_UNCOVER_2025}.
Multiple images of the same galaxy have been removed. Galaxy positions are de-lensed to the source plane with the lens model of Section~\ref{sec:lens_model}. Fig.~\ref{fig:environment_density} shows the projected source-plane distance of LAP2 from its neighbors, color-coded by their redshift offset. At $z=4.2$, $|\Delta z|=0.05$ corresponds to $\Delta v\simeq2900$~km\,s$^{-1}$, or $\simeq6$~pMpc along the line of sight. 
%, and 1~pMpc corresponds to $\simeq2.4$~arcmin. 
Only 5 spectroscopically confirmed galaxies are found within this redshift range, in addition to LAP2; the projected closest neighbor to LAP2 in the source plane is found at a distance of $\rm \sim80~pkpc$. More generally, LAP2 has only one spectroscopic neighbor within a cylinder of $\rm R<0.5$~pMpc in radius and $\rm |\Delta d_{\rm LOS}|<2$~pMpc along the line of sight; only $\rm 11\%$ of the sources in the spectroscopic catalog have as few neighbors within the same cylinder. 
Moreover, LAP2 lies in a near-empty redshift interval ($4.06\lesssim z\lesssim4.26$) between two concentrations of galaxies at $z\simeq3.96$--$4.04$ and $z\simeq4.28$--$4.32$ (see bottom panel of Fig.~\ref{fig:environment_density}).

This is in line with the other extremely metal-poor candidates found at intermediate redshift: CR3 at $z=3.19$ lies in an average-to-slightly-underdense region \citep[$\delta\simeq-0.12$;][]{Cai2025}, and AMORE6 at $z=5.73$ sits in the galaxy-poor gap between two overdensities \citep{Morishita2025}. A low-density environment slows halo assembly and keeps a system far from neighbors that could pre-enrich it through outflows (which reach $\lesssim$100--200~pkpc) or disturb it with their radiation. This lets pockets of nearly pristine gas survive well after reionization, as predicted by models of late Pop~III star formation \citep[e.g.,][]{Tornatore2007,Trenti2009b,Xu2016,Venditti2023}. 
LAP2 appears to lie in a sparsely populated region, with no evident massive or star-forming neighbors. A formal overdensity estimate is not possible, given the inhomogeneous spectroscopic selection and the reduced source-plane volume behind the cluster; nevertheless, the isolation of LAP2 supports interpreting it as a site of (nearly) pristine star formation at $z\sim4$.

Low-density regions can delay the assembly and enrichment of such systems, but not indefinitely: as structure grows, even mildly under-dense environments are eventually incorporated into larger structures. Simulations (and observations) suggest that, by $z=0$, the remnants of late, pristine star formation end up predominantly in the stellar halos and low-mass satellites of more massive galaxies \citep[e.g.,][]{Pustilnik2019,Storck2026}.

\section{Conclusions} \label{sec:conclusions}
We have presented new JWST/NIRSpec IFU PRISM observations (GO~7677) of LAP2, a strongly lensed galaxy at $z=4.2$ behind Abell~2744. We combined them with NIRCam imaging to study its morphology, stellar population, ionizing properties, and oxygen content in a spatially resolved way. Our main results are the following:

\begin{itemize}
\item LAP2a, the most magnified counter-image ($\mu_{\rm tot}=51.3$, $\mu_{\rm tan}=21.9$), consists of a compact core ($R_{\rm eff}\simeq7$~pc), embedded in fainter emission with $R_{\rm eff}\simeq20$~pc. With $M_{\rm UV}=-12.7$ and $\log(M_\star/M_\odot)\simeq4.9$, LAP2 is among the faintest and least massive galaxies with spectroscopic metallicity constraints at high redshift, and its mass is comparable to that of massive star clusters.
 
\item The system hosts a young, dust-free starburst: H$\alpha$ is detected with EW$_{\rm rest}\simeq880$~\AA\ and $\log(\xi_{\rm ion}/{\rm erg^{-1}\,Hz})\simeq25.3$, the Balmer decrement (H$\alpha$/H$\beta=2.5\pm0.5$) is consistent with no attenuation, and SED fitting including the emission lines returns a mass-weighted age $\lesssim10$~Myr. The blue UV slope ($\beta_{\rm UV}\simeq-2.5$, down to $\simeq-3.1$ in the core) and the underprediction of the bluest bands by models without escape suggest a substantial escape of ionizing photons, although this remains tentative given the photometric uncertainties.
 
\item {[}O\,{\sc iii}]$\lambda$5007 is not detected in the integrated spectrum of LAP2a, giving [O\,{\sc iii}]/H$\beta\leq0.61$ ($3\sigma$). With the \citet{Sanders2024} calibration, extrapolated to this regime, this corresponds to $12+\log({\rm O/H})\leq6.4$ ($\lesssim0.5\%\,Z_\odot$). This confirms, on an independent counter-image and with deeper data, the oxygen deficiency first suggested by the NIRSpec/MSA spectrum of LAP2b \citep{Vanzella2026b}. The large
EW(H$\alpha$) and $\xi_{\rm ion}$ disfavor a low ionization explanation for the weak [O\,{\sc iii}]. The non-detection of He\,{\sc ii}$\lambda$1640 (EW$_{\rm rest}<39$~\AA, $3\sigma$) rules out only the most extreme, zero-age metal-free bursts, but does not exclude a contribution from Pop~III stars mixed with the extremely metal-poor population.
 
\item The IFU data reveal that the oxygen deficiency is not spatially uniform. Faint [O\,{\sc iii}] emission is detected in the compact core, where we derive $12+\log({\rm O/H})\simeq6.6$ ($\rm \sim1\%~Z_\odot$). The difference can be explained either by a higher ionization parameter in the denser core or by local self-enrichment. The latter requires only $\rm \sim0.2\,M_\odot$ of newly synthesized oxygen, i.e. the ejecta of $\sim0.1\%$ of the stellar mass formed in the core.
  
\item We identify a tentative H$\alpha$ component $\gtrsim30$~pc north-east of the core, with no stellar-continuum counterpart and a velocity offset of $\sim400$~km\,s$^{-1}$. If confirmed, it could trace gas ionized by radiation escaping from the core or a very young stellar clump below the NIRCam detection limit.
 
\item Compared with other faint high-redshift galaxies, the integrated [O\,{\sc iii}]/H$\beta$ limit of LAP2 lies $\gtrsim0.3$~dex below the plateau of the $R3$--$M_{\rm UV}$ relation of \citet{Korber2026}, extrapolated to $M_{\rm UV}\approx-13$. In the mass--metallicity plane, LAP2 is one of the most metal-poor systems known and one of the few with a star-cluster-like stellar mass. No clear redshift dependence emerges among the known EMPGs, which are found from $z\approx3$ to $z>10$, likely because of the limited sample size.
 
\item LAP2 lies in a sparsely populated region: it has only one spectroscopically confirmed neighbor within 0.5~projected pMpc and within $\pm2$~pMpc along the line of sight ($|\Delta v|\lesssim450$~km\,s$^{-1}$), fewer than $\sim$90\% of galaxies at similar redshift in the same field. It also sits in a near-empty redshift interval between two galaxy concentrations at $z\simeq4.0$ and $z\simeq4.3$. This configuration resembles that of other EMPGs at intermediate redshift (CR3, AMORE6), and supports a scenario in  which low-density environments preserve nearly pristine gas well after reionization.
\end{itemize}

Beyond confirming the oxygen deficiency of LAP2 on an independent counter-image, the IFU observations reveal that its chemical state is not uniform: oxygen is detected only in the compact core, while the rest of the system remains undetected in [O\,{\sc iii}]. This shows that spatially resolved spectroscopy is required to identify the most pristine regions of EMPGs and Pop~III candidates, which may coexist with, or lie close to, more enriched components on scales of tens to hundreds of pc. The number of such candidates is expected to grow quickly, as photometric selections based on colour--colour diagnostics are now being used to identify them for spectroscopic follow-up \citep[e.g.][]{Trussler2023,Trussler2026,Fujimoto2025_arxiv,Fujimoto2025}; although not free from contaminants \citep[e.g.][]{Fujimoto2025_arxiv}, these selections will help populate the low-mass, low-metallicity end of the MZR. 
%For LAP2, upcoming deeper NIRSpec/MSA spectroscopy will search for fainter diagnostic lines, while the combination with the spatial information provided by the IFU will be needed to test whether metal-free stars contribute to its emission.

\begin{acknowledgements}
This work was supported by the INAF Excellence Track grant Astrofisica di Frontiera 2025, PARSE "PARsec-scale Star formation in the Early Universe: from VLT/JWST to the ELT" (PI Vanzella). MM and EV acknowledge financial support through INAF GO Grant 2024 ``Mapping Star Cluster Feedback in a Galaxy 450 Myr after the Big Bang'' and the project PRORIS - COSMOWEB ``A new era for cosmology:  exploiting the JWST revolution''. MC acknowledges INAF GO Grant 2024 ”Revealing the nature of bright galaxies at cosmic dawn with deep JWST spectroscopy”.  
Support for SRR and TT under program JWST-GO-7677 was provided by NASA through a grant from the Space Telescope Science Institute, which is operated by the Association of Universities for Research in Astronomy, Inc., under NASA contract NAS5-03127.
Part of the data products presented herein were retrieved from the Dawn JWST Archive (DJA). DJA is an initiative of the Cosmic Dawn Center (DAWN), which is funded by the Danish National Research Foundation under grant DNRF140.
This research has used NASA’s Astrophysics Data System, QFitsView, and SAOImageDS9, developed by Smithsonian Astrophysical Observatory.
Additionally, this work made use of the following open-source packages for Python, and we are thankful to the developers of these: Matplotlib \citep{matplotlib2007}, Numpy \citep[][]{NUMPY2011}, Astropy \citep{astropy:2022} (http://www.astropy.org).
\end{acknowledgements}

% WARNING
%-------------------------------------------------------------------
% Please note that we have included the references to the file aa.dem in
% order to compile it, but we ask you to:
%
% - use BibTeX with the regular commands:
%   \bibliographystyle{aa} % style aa.bst
%   \bibliography{Yourfile} % your references Yourfile.bib
%
% - join the .bib files when you upload your source files
%-------------------------------------------------------------------

\bibliographystyle{aa}
\bibliography{bib}

\begin{appendix} %First appendix
\onecolumn
\section{LAP2a analysis on the 100 mas/pixel reduction}\label{sec:app:spec_further}
\begin{figure*}[h]
    \centering
    \includegraphics[width=0.99\textwidth]{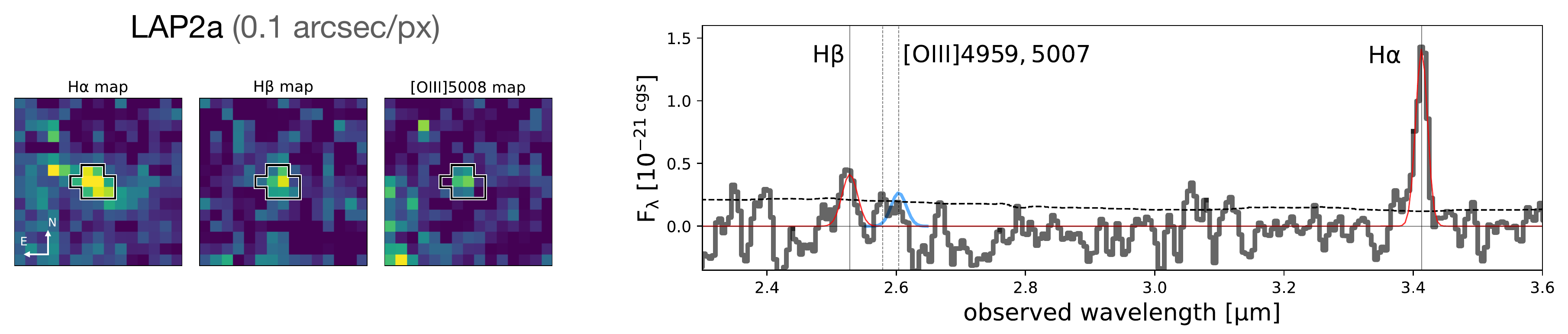}
    \caption{Emission line maps of \ha, \hb\ and \oiiiv\ (left panels) and extracted spectrum (right panel) for LAP2a, in the data reduction test at 0''.1 pixel scale. The same color-coding as of Fig.~\ref{fig:spec_main} in the main text is used.}
    \label{fig:app:LAP2a_pxsc01}
\end{figure*}

%\subsection{Tests on single pointings}\label{sec:app:single_pointing}
%\subsection{Tests on background subtractions}\label{sec:app:bkg}
%\subsection{Tests on the 100 mas/pixel cube}\label{sec:app:100mas}
Fig.~\ref{fig:app:LAP2a_pxsc01} collects the emission line maps and the spectrum of LAP2a from a reduction at $\rm 0.1$ arcsec/pixel. This side-test confirms the main results found in the text: we detect both \ha\ and \hb\ emission, while a possible \oiiiv\ sits below the detection limit, giving a $\rm 3\sigma$ upper limit $\rm R3\leq0.62$. Also at this pixel scale a tentative \oiiiv\ emission is detected in few pixels; the derived $\rm R3=0.64^{+0.33}_{-0.23}$ is consistent with the value found in the main analysis ($\rm R3=1.11\pm0.24$). 

\section{Analysis of LAP2b}\label{sec:app:lap2b}
\begin{figure*}[h]
    \centering
    \includegraphics[width=0.99\textwidth]{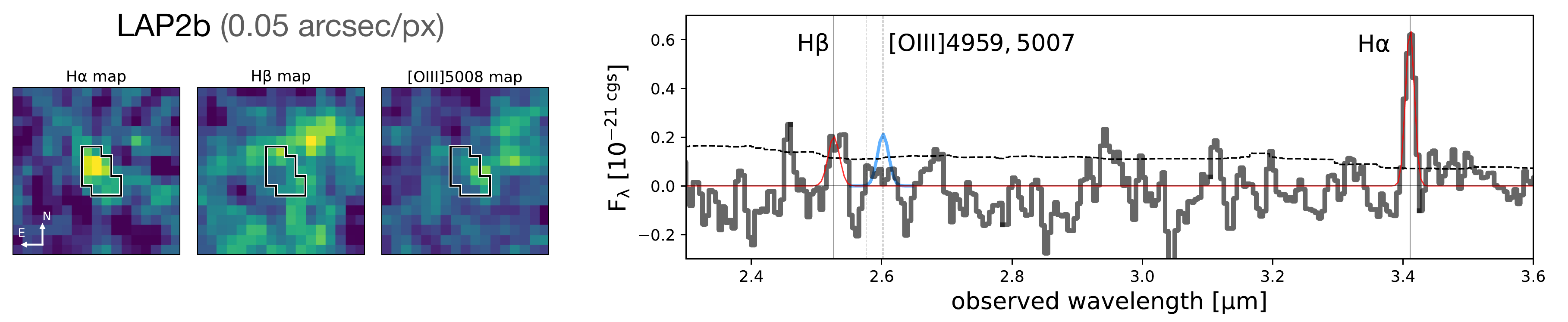}
    \caption{Emission line maps of \ha, \hb\ and \oiiiv\ (left panels) and extracted spectrum (right panel) for the LAP2b image. The same color-coding as of Fig.~\ref{fig:spec_main} in the main text is used.}
    \label{fig:app:LAP2b}
\end{figure*}

From the spectrum of counterimage LAP2b (Fig.~\ref{fig:app:LAP2b}) we detect both \ha\ ($\rm \mu F_{H\alpha}=1.09^{+0.19}_{-0.20}\cdot10^{-19}~[erg~s^{-1} cm^{-2}]$) and \hb\ ($\rm \mu F_{H\beta}=4.50^{+1.45}_{-1.42}\cdot10^{-20}~[erg~s^{-1} cm^{-2}]$), the latter with low SNR; no \oiii\ line is detected, resulting in a $\rm 3\sigma$ upper limit on \oiiiv\ $\rm<4.56\cdot10^{-20}~[erg~s^{-1} cm^{-2}]$ ($\rm R3\lesssim1$), less constraining that the value derived for LAP2a.  
We point out that LAP2b sits close to a lower redshift galaxy (only $\sim0''.15$ away) and that in our NIRSpec-IFS observations it is covered only by one of the two pointings of the mosaic, sitting near the edge of the field-of-view. These factors explain the higher noise observed in this region (see emission-line maps in Fig.~\ref{fig:app:LAP2b}). A possible faint \oiiiv\ emission appears in the map, but it appear shifted from the peak of the \ha\ emission. Because of the low SNR reached in this region we preferred to exclude LAP2b from the main analysis of the paper. The rest-UV flux of LAP2b (derived from photometry) is $\sim1.2\times$ lower than for LAP2a, consistent with the magnification ratio between the two counterimates, $\rm \mu_{LAP2a}/\mu_{LAP2b}=1.09\pm0.12$ (Section~\ref{sec:lens_model}); the higher ratios derived for the \ha\ and \hb\ fluxes (a factor $\sim3$) should be attributed to the small mask used to extract the spectrum of LAP2b, optimized not to include noisy regions. 
\end{appendix}

\end{document}